\documentclass[a4paper,aps,prd,preprint,tightenline]{revtex4-1}

\usepackage{color}
\usepackage{amsmath}
\usepackage{accents}
\usepackage{amssymb}
\usepackage{graphicx}
\usepackage{marginnote}

\usepackage[colorlinks]{hyperref} 
\hypersetup{
	citecolor=blue,
	linkcolor=blue,
	urlcolor=blue
}

\newcommand{\defineText}[2]{\newcommand{#1}[0]{\text{#2}}}

\newlength{\dhatheight}

\defineText{\MV}{MV}
\defineText{\GV}{GV}
\defineText{\MeV}{MeV}
\defineText{\GeV}{GeV}
\defineText{\TeV}{TeV}
\defineText{\cm}{cm}
\defineText{\km}{km}
\defineText{\kpc}{kpc}
\defineText{\sr}{sr}
\defineText{\LIS}{LIS}
\defineText{\TOA}{TOA}

\usepackage{xcolor}

\makeatother

\usepackage{babel}

\usepackage{comment}

\renewcommand{\figurename}{FIG.}

\renewcommand{\tablename}{TAB.} 

\newcommand{\beq}[1]{\begin{equation}\label{#1}}
	\newcommand{\eeq}{\end{equation}}
\newcommand{\bea}[1]{\begin{eqnarray}\label{#1}}
	\newcommand{\eea}{\end{eqnarray}}
	
\usepackage{tcolorbox}
\newcounter{reviewer}
\newcounter{point}[reviewer]
\renewcommand{\thepoint}{\thereviewer.\arabic{point}}

\begin{document}
\title{Constraints on Primordial Black Holes from Galactic Diffuse Synchrotron Emissions}
\author{Chen-Wei Du}
\affiliation{Institute of Theoretical Physics, 
	Chinese Academy of Sciences, Beijing 100190, China. \\
	University of Chinese Academy of Sciences, Beijing, 100190, China.}
\author{Yu-Feng Zhou}
\affiliation{Institute of Theoretical Physics, 
	Chinese Academy of Sciences, Beijing 100190, China. \\
	University of Chinese Academy of Sciences, Beijing, 100190, China.}
\affiliation{School of Fundamental Physics and Mathematical Sciences, 
	Hangzhou Institute for Advanced Study, UCAS, Hangzhou 310024, China. \\
	International Centre for Theoretical Physics Asia-Pacific, Beijing/Hangzhou,
China.}
\date{\today}
\begin{abstract} 
We investigate the possibility of constraining primordial black holes (PBHs) with masses $M_\mathrm{PBH}\gtrsim 10^{15}\,\mathrm{g}$ through Galactic diffuse synchrotron emissions. 
Due to Hawking radiation, these types of PBHs are expected to be stable sources of cosmic-ray (CR) electrons and positrons with energies below $\mathcal{O}(10\,\mathrm{MeV})$. 
In many CR propagation models with diffusive re-acceleration characterized by a significant Alfv\'{e}n velocity $V_a\sim \mathcal{O}(10)\,\mathrm{km/s}$, the energies of the evaporated electrons/positrons can be further enhanced to $\mathcal{O}(100)\,\mathrm{MeV}$ through their scattering with the Galactic random magnetic fields. 
Consequently, the observation of Galactic synchrotron emissions at frequencies above $\sim 20\,\mathrm{MHz}$ can provide useful constraints on the abundance of PBHs. 
Using the AMS-02 and Voyager-1 data on the boron-to-carbon nuclei flux ratio, we confirm that a significant Alfv\'{e}n velocity $V_a \sim 20\,\mathrm{km/s}$ is favored in several benchmark diffusive re-acceleration models. 
We show that, in this scenario, the observed low-frequency synchrotron emissions (from 22 MHz to 1.4 GHz) can provide stringent constraints on PBH abundance. 
The obtained conservative constraints are stronger than those derived from the Voyager-1 all-electron (electron plus positron) data by more than one order of magnitude for
$M_\mathrm{PBH}\gtrsim 1\times 10^{16}\,\mathrm{g}$, and also stronger than our previous constraints derived from the AMS-02 positron data for $M_\mathrm{PBH}\gtrsim 2\times 10^{16}\,\mathrm{g}$. 
\end{abstract}
\maketitle

\section{Introduction}
Several astrophysical and cosmological evidences point towards the existence of dark matter (DM), something that constitutes $\sim 26\%$ of the total energy density of the Universe \cite{Planck:2018vyg}. 
Despite its abundance, the nature of DM remains mysterious, as it has evaded all nongravitational direct and indirect detections thus far \cite{Bertone:2004pz, Slatyer:2017sev, Lin:2019uvt}. 
One of the most widely discussed DM candidates is primordial black hole (PBH) \cite{Hawking:1971ei, Carr:1974nx, Chapline:1975ojl, Meszaros:1975ef, Carr:1975qj}. 
PBHs may have formed in the early universe from the collapse of overdensities resulting from quantum fluctuations \cite{Tashiro:2008sf, Germani:2018jgr, Kannike:2017bxn, Carr:2017jsz, Carr:2017edp} and by other mechanisms such as phase transitions \cite{Crawford:1982yz, Hawking:1982ga, Khlopov:2008qy, Belotsky:2014kca, Byrnes:2018clq, Kitajima:2020kig, Dvali:2021byy, Khlopov:2024nqp}. 
The mass of PBHs can vary in a large range depending on the formation time. In general, the initial mass $M_\mathrm{PBH}$ of a PBH should be close to the mass enclosed by the Hubble horizon at the formation time $t$, i.e., $M_\mathrm{PBH} \sim c^3 t / G \simeq (t / 10^{-23}\,\mathrm{s}) \, 10^{15}\,\mathrm{g}$, where $c$ is the speed of light and $G$ is the Newton constant. For two typical formation times: the Planck time ($t \sim 10^{-43}\,\mathrm{s}$) and the time just before the big-bang nucleosynthesis ($t \sim 1\,\mathrm{s}$), the initial PBH masses are approximately $10^{-5}\,\mathrm{g}$ and $10^{5}\,M_\odot$, respectively, where $M_\odot$ is the solar mass. 
Moreover, realistic production mechanisms predict not just a unique mass for all PBHs but rather an extended mass function. 

PBHs may constitute all or a fraction of the DM. The fraction of DM in the form of PBHs is defined as $f_\mathrm{PBH} \equiv \Omega_\mathrm{PBH} / \Omega_\mathrm{DM}$, where $\Omega_\mathrm{PBH}$ and $\Omega_\mathrm{DM}$ are the energy density parameters of PBHs and DM relative to the critical density of the present Universe, respectively. 
There exist numerous observational constraints on $f_\mathrm{PBH}$ (for recent reviews, see, e.g., \cite{Belotsky:2018wph, Carr:2020gox, Carr:2021bzv, Auffinger:2022khh}). For heavy PBHs with masses $M_\mathrm{PBH} \gg 10^{17}\,\mathrm{g}$, constraints on $f_\mathrm{PBH}$ arise from gravitational effects of PBHs, including lensing \cite{MACHO:2000qbb, Macho:2000nvd, Wyrzykowski:2010mh, Wyrzykowski:2011tr, Griest:2013aaa, CalchiNovati:2013jpj, PhysRevLett.111.181302}, gravitational waves \cite{Raidal:2017mfl, Raidal:2018bbj, Vaskonen:2019jpv, Chen:2019xse, Saito:2008jc, Assadullahi:2009jc, Bugaev:2010bb, Hong:2026rcl}, dynamical constraints from globular clusters, Galaxy disruption, and other observables \cite{Carr:1997cn, Monroy-Rodriguez:2014ula, Brandt:2016aco, Koushiappas:2017chw, Carr:2018rid}. 
Light PBHs are expected to emit Standard Model particles with a quasi-thermal energy spectrum through Hawking radiation \cite{Hawking:1974rv, Hawking:1975vcx}, where the temperature of the Hawking spectrum is inversely proportional to the PBH mass. 
Due to Hawking radiation, PBHs lose their mass at a rate $\mathrm{d}M_\mathrm{PBH}/\mathrm{d}t \propto M_\mathrm{PBH}^{-2}$ \cite{Baldes:2020nuv, Cheek:2021odj}, implying that lighter PBHs evaporate more quickly. It has been shown that PBHs with masses $M_\mathrm{PBH}<5\times 10^{14}\,\mathrm{g}$ have lifetimes shorter than the age of the Universe \cite{MacGibbon:1991tj}, thus cannot contribute to present DM. 
In this work, we are particularly interested in the mass range above the evaporation limit ($5\times 10^{14}\,\mathrm{g}$) and below the lowest lensing limit ($5\times 10^{17}\,\mathrm{g}$). 
In this mass range, PBHs are expected to emit particles with typical energies from $\mathcal{O}(10\,\mathrm{MeV})$ down to $\mathcal{O}(10\,\mathrm{keV})$. 
A number of constraints on $f_\mathrm{PBH}$ have been set by considering Hawking radiation in various astrophysical observations, such as extragalactic and galactic $\gamma$-rays \cite{Carr:2009jm, Carr:2016hva, Arbey:2019vqx, Laha:2020ivk, Ray:2021mxu}, the 511 keV line from the Galactic Center \cite{Dasgupta:2019cae, Laha:2019ssq, Keith:2021guq, DelaTorreLuque:2024qms}, cosmic microwave background (CMB) \cite{Clark:2016nst, Acharya:2020jbv, Auffinger:2022khh}, 21cm radio signals \cite{Mittal:2021egv, Saha:2021pqf}, Lyman-$\alpha$ forest measurements \cite{Saha:2024ies, Khan:2025kag}, neutrinos \cite{Dasgupta:2019cae, Wang:2020uvi, Bernal:2022swt, Liu:2023cqs}, cosmic-ray (CR) electrons \cite{Boudaud:2018hqb}, CR positrons \cite{Huang:2024xap} and CR antiprotons \cite{Maki:1995pa, Barrau:2001ev}. 

Recently, it has been noted that strong constraints on $f_\mathrm{PBH}$ can be derived from direct and indirect CR electron and positron observables when adopting diffusive re-acceleration models of CR propagation \cite{Huang:2024xap, DelaTorreLuque:2024qms}. 
Such models, characterized by a significant Alfv\'{e}n velocity $V_a\sim\mathcal{O}(10)\,\mathrm{km/s}$, have received strong support from a number of independent analyses \cite{Trotta:2010mx, Jin:2014ica, Johannesson:2016rlh, Boschini:2017fxq, Boschini:2018baj, Boschini:2020jty, DeLaTorreLuque:2021yfq, Luque:2021nxb, Yuan:2017ozr, Korsmeier:2021brc, Silver:2024ero}. 
In these models, the energies of the evaporated all-electrons (electrons plus positrons), which typically have initial energies of $\mathcal{O}(10\,\mathrm{MeV})$, can be enhanced by roughly two orders of magnitude through scattering with the Galactic random magnetic fields during propagation. 
This effect makes it possible to constrain $f_\mathrm{PBH}$ using direct and indirect CR electron and positron observables at energies around the GeV scale. 
For instance, our previous work \cite{Huang:2024xap} showed that, in well-constrained diffusive re-acceleration models, a significant portion of the flux of evaporated positrons could be constrained by current AMS-02 data, deriving constraints approximately an order of magnitude stronger than those derived from Voyager-1 all-electron data for $M_\mathrm{PBH}\sim 10^{16}\,\mathrm{g}$. 
Similarly, in \cite{DelaTorreLuque:2024qms}, it was shown that the diffuse X-ray emissions from the up-scattering of Galactic ambient photons due to the inverse Compton effect of the evaporated all-electrons could be constrained by the XMM-Newton data, and the obtained constraints could be improved by two orders of magnitude for $M_\mathrm{PBH}\sim 10^{16}\,\mathrm{g}$ as $V_a$ increasing from 13.4 km/s to 40 km/s in their benchmark diffusive re-acceleration models. 
Motivated by these works, we investigate another possible indirect detection channel of PBHs -- Galactic diffuse synchrotron emission. 
The Galactic diffuse synchrotron emission arises from CR all-electrons during their propagation in the Galactic magnetic field (GMF), and constitutes an indirect observable of the interstellar CR all-electrons. 
The observation of Galactic synchrotron emissions at frequencies above $\sim20\,\mathrm{MHz}$ can provide indirect measurements of interstellar CR all-electrons with energies above $\sim 100\,\mathrm{MeV}$, and therefore can set meaningful constraints on $f_\mathrm{PBH}$ within diffusive re-acceleration models. 

To discuss the impact of diffusive re-acceleration on the constraints derived from Galactic synchrotron observations, we fit parameters of a set of benchmark CR propagation models to the AMS-02 \cite{AMS:2023anq} and Voyager-1 \cite{Cummings:2016pdr} data on the boron-to-carbon nuclei (B/C) flux ratio. 
Our model set includes four diffusive re-acceleration models with different diffusion halo half-heights $z_\mathrm{h}$ (fixed to 4, 6, 8, 10 kpc, respectively), and one diffusion break model with $z_\mathrm{h}$ fixed to 4 kpc, which does not include the diffusive re-acceleration (i.e., with $V_a$ fixed to zero). 
For diffusive re-acceleration models, our fits obtain significant Alfv\'{e}n velocities of $V_a \sim 20\,\mathrm{km/s}$. 
We also employ an additional diffusive re-acceleration model with the parameters fitted in \cite{Boschini:2020jty}, in which solar modulation is modeled using the \texttt{Helmod} code \cite{Bobik:2011ig, bobik2016forward, Boschini:2017gic, Boschini:2018zdv, Boschini:2019ubh, Boschini:2022jwz}, to complement our simple treatment of force-field approximation. 
We show that, for $V_a\sim 20\,\mathrm{km/s}$, a significant fraction of the evaporated all-electrons can be boosted to the energies of $\sim 100\,\mathrm{MeV}$ during their Galactic propagation. 
Consequently, synchrotron emissions generated by the evaporated all-electrons, hereafter referred to as synchrotron signals of PBHs, can be constrained by the low-frequency radio continuum surveys (from 22 MHz to 1.4 GHz). 
With diffusive re-acceleration models, we obtain stringent constraints on $f_\mathrm{PBH}$. 
The most conservative constraints are stronger than those derived from the Voyager-1 all-electron data \cite{Boudaud:2018hqb} by more than one order of magnitude for $M_\mathrm{PBH} \gtrsim  1\times 10^{16}\,\mathrm{g}$, and also stronger than our previous constraints derived from the AMS-02 positron data \cite{Huang:2024xap} for $M_\mathrm{PBH} \gtrsim 2\times 10^{16}\,\mathrm{g}$. 

The remainder of this paper is organized as follows. 
In section~\ref{sec: PBH evaporation}, we provide a brief overview of the all-electron energy spectrum from PBH evaporation. 
In section~\ref{sec: CR}, we describe the CR propagation in the Galaxy and the models for solar modulation, and present the parameter fit results for our benchmark CR propagation models. 
In section~\ref{sec: synch and GMF}, we describe the Galactic synchrotron emission and the adopted GMF models. 
In section~\ref{sec: fPBH}, we discuss the constraints on $f_\mathrm{PBH}$ derived from Galactic synchrotron emission observations. 
Finally, we summarize our work in section~\ref{sec: conclusion}.

\section{Evaporation of Primordial Black Holes}   \label{sec: PBH evaporation}
In this section, we adopt the natural system of units with $\hbar=k_\mathrm{B}=c=1$, where $\hbar$ is the reduced Planck constant, $k_\mathrm{B}$ is the Boltzmann constant, and $c$ is the speed of light. 
We consider a simple scenario that the spin of PBHs is negligible, which can be obtained from a series of possible formation mechanisms \cite{DeLuca:2019buf, Chiba:2017rvs, Mirbabayi:2019uph}. 
The emission rate of particle species $i$ per unit total energy $E$ from a PBH of mass $M_{\mathrm{PBH}}$ is given by \cite{Hawking:1975vcx} 
\begin{equation}
    \frac{\mathrm{d}^2 N_i}{\mathrm{d}t\mathrm{d}E} = \frac{g_i \Gamma_i}{2\pi}\left[\exp\left(\frac{E}{T_\mathrm{PBH}}\right) - (-1)^{2s_i} \right]^{-1} \ ,   \label{eq:Hawking radiation}
\end{equation}
where $s_i$ and $g_i$ are the spin and the total degree of freedom of the particle $i$, respectively, $\Gamma_i$ is the energy-dependent graybody factor of the particle $i$, and $T_\mathrm{PBH}$ is the temperature of the PBH which is given by \cite{Page:1976df} 
\begin{equation}
    T_\mathrm{PBH} \approx 10.6\times \left(\frac{10^{15}\,\mathrm{g}}{M_\mathrm{PBH}}\right) \,\mathrm{MeV} \ .
\end{equation}
The graybody factor $\Gamma_i$ in Eq.~\eqref{eq:Hawking radiation} describes the probability that the particle $i$ created at the PBH horizon finally escapes to spatial infinity, and is determined by the equation of motion of the particle in curved spacetime near the horizon. In the geometric optics limit (i.e., the high-energy limit), the graybody factor for electrons can be approximated as $\Gamma_e \simeq 27G^2 M_\mathrm{PBH}^2 E^2$. 
Note that Eq.~\eqref{eq:Hawking radiation} only describes the primary particles directly emitted from the PBH. The production of secondary particles from decays of unstable primary particles should also be considered. We use the numerical code \texttt{BlackHawk} \cite{Arbey:2019mbc, Arbey:2021mbl} to compute the energy spectra of evaporated all-electrons, in which both the primary and secondary production processes are calculated. Following \texttt{BlackHawk}'s manual, we calculate the secondary production process using the results of \texttt{Hazma} code \cite{Coogan:2019qpu}, since the energies of primary particles cannot reach 5 GeV for PBH masses we interested in. 

The mass function of PBHs, $\mathrm{d}n/\mathrm{d}M_{\mathrm{PBH}}$ (number density per unit mass), depends on the formation mechanisms of PBHs. 
In this work, we consider two widely used mass functions: monochromatic \cite{Carr:2017jsz} and log-normal \cite{Kannike:2017bxn} mass functions. A nearly monochromatic mass function is expected if all PBHs are formed at a same epoch, and a log-normal mass function can arise from inflationary fluctuations \cite{Dolgov:1992pu, Clesse:2015wea}. These two mass functions are given by 
\begin{equation}
    \frac{\mathrm{d}n}{\mathrm{d}M_{\mathrm{PBH}}} = \left\{ \begin{aligned} & A_1 \delta(M_{\mathrm{PBH}}-M_{\mathrm{c}}) & \text{(monochromatic)} \\ & \frac{A_2}{\sqrt{2\pi}\sigma M_\mathrm{PBH}^2} \mathrm{exp}\left[ - \frac{\mathrm{ln}^2(M_\mathrm{PBH}/M_\mathrm{c})}{2\sigma^2} \right] & \text{(log-normal)} \ ,  \end{aligned} \right.  
    \label{eq:PBH mass distribution}
\end{equation}
where $M_\mathrm{c}$ is the characteristic mass, $\sigma$ is the width of log-normal mass distribution, and $A_1$ and $A_2$ are normalization factors determined by the energy density of PBHs, $\rho_\mathrm{PBH}$, as follows: 
\begin{equation}
    \int^\infty_{M_\mathrm{min}} M_\mathrm{PBH} \frac{\mathrm{d}n}{\mathrm{d}M_\mathrm{PBH}} \mathrm{d}M_\mathrm{PBH} = \rho_\mathrm{PBH} \ , 
    \label{eq:PBH norm}
\end{equation}
where $M_\mathrm{min} = 5 \times 10^{14}\,\mathrm{g}$ corresponds to the mass of which PBHs have completely evaporated today \cite{MacGibbon:1991tj}, thus the PBHs lighter than $M_\mathrm{min}$ cannot contribute to present DM. For PBHs heavier than $M_\mathrm{min}$, the evaporation timescales drastically exceed the timescales of CR propagation in the Galaxy, so we neglect the variation of $M_\mathrm{PBH}$ during evaporation. 

The energy spectrum of all-electrons evaporated by PBHs with a mass function $\mathrm{d}n/\mathrm{d}M_{\mathrm{PBH}}$ is given by 
\begin{equation}
    \frac{\mathrm{d}^2 n_{e^\pm}}{\mathrm{d}t\mathrm{d}E} = \int^\infty_{M_\mathrm{min}} \frac{\mathrm{d}n}{\mathrm{d}M_{\mathrm{PBH}}} \frac{\mathrm{d}^2 N_{e^\pm}}{\mathrm{d}t\mathrm{d}E} \,\mathrm{d}M_\mathrm{PBH} \ ,
\end{equation}
where $\mathrm{d}^2 N_{e^\pm}/\mathrm{d}t\mathrm{d}E$ is the emission rate of all-electrons from a single PBH of mass $M_\mathrm{PBH}$ with both the primary and secondary emission components considered. 
For the monochromatic mass function, above $M_\mathrm{PBH}$ integral simplifies easily. For the log-normal mass function, the integral is calculated by \texttt{BlackHawk}.

\section{Cosmic Ray Propagation}  \label{sec: CR}
\subsection{Cosmic Ray Propagation in the Galaxy}   \label{subsec: CR propagation}
Once injected into the Galaxy, the all-electrons evaporated by PBHs propagate as CRs. 
When discussing CR propagation, we assume cylindrical symmetry of the Galaxy, and use cylindrical coordinates $(R,\, z)$. We assume that CR propagation occurs within a cylindrical diffusion halo of radius $R_\mathrm{h} = 20\,\mathrm{kpc}$ and half-height $z_\mathrm{h}$ (typically a few kpc). The number density of CR particles per unit momentum $p$ at position $\vec{r}$ at time $t$, $\psi(\vec{r}, p, t)$, is related to the phase space distribution function $f(\vec{r},\vec{p},t)$ as $\psi(\vec{r}, p, t)=4\pi p^2f(\vec{r},\vec{p},t)$, assuming an isotropic momentum distribution of CR particles. A free-escape boundary condition ($\psi = 0$) is imposed at the boundary of diffusion halo. The Galactic propagation of CRs is described by the diffusion equation: 
\begin{equation}
    \begin{aligned}
    \frac{\partial \psi}{\partial t} =& q(\vec{r}, p)+ \nabla \cdot\left(D_{x x} \nabla \psi-\vec{V}_c \psi\right)+\frac{\partial}{\partial p} p^2 D_{p p} \frac{\partial}{\partial p} \frac{1}{p^2} \psi \\
    & -\frac{\partial}{\partial p}\left[\dot{p} \psi-\frac{p}{3}(\nabla \cdot \vec{V}_c) \psi\right]-\frac{1}{\tau_f} \psi-\frac{1}{\tau_r} \psi \ ,
    \end{aligned}
    \label{eq: prop eq}
\end{equation}
where $q$ is the source term, which represents the number of CR particles injected into the Galaxy per unit volume, per unit momentum, and per unit time, $D_{xx}$ is the spatial diffusion coefficient, $\vec{V}_c = \mathrm{sign}(z) \left( V_0 + \frac{\mathrm{d}V}{\mathrm{d}z} |z| \right) \hat{z}$ is the convection velocity driven by the galactic wind with $V_0$ generally assumed zero, $D_{pp}$ is the diffusion coefficient in momentum space, describing the effect of diffusive re-acceleration, $\dot{p}\equiv\mathrm{d}p/\mathrm{d}t$ is the momentum loss rate, which accounts for physical processes through which CR particles lose energy during propagation, including ionization, Coulomb interactions, bremsstrahlung, inverse Compton and synchrotron processes, $\tau_f$ and $\tau_r$ are the timescales of particle fragmentation and radioactive decay, which account for destruction of CR particles by interaction with interstellar gas and decay of unstable CR particles, respectively. 
We employ the state-of-the-art \texttt{Galprop} code \cite{Strong:1998pw, Moskalenko:2001ya, Strong:2001fu, Moskalenko:2002yx, Strong:2007nh} to solve the diffusion equation numerically. 
Following the Eq.~\eqref{eq: prop eq}, \texttt{Galprop} evolves $\psi$ long enough time to make $\partial\psi/\partial t = 0$ satisfied, and the result is the steady-state solution of the diffusion equation. 

In the diffusion equation Eq.~\eqref{eq: prop eq}, the spatial diffusion coefficient $D_{xx}$ is parameterized as follows: 
\begin{equation}
    D_{xx} = D_0 \beta^\eta \left( \frac{\rho}{\rho_0} \right)^\delta \ ,
    \label{eq:diff_coeff}
\end{equation}
where $\rho$ is the rigidity of CR particles, $\beta = v/c$ is the velocity of CR particles relative to the speed of light $c$, $D_0$ is the normalization of diffusion coefficient at reference rigidity $\rho_0$, and $\delta$ is the spectral power index, for Kolmogorov type of turbulence $\delta = 1/3$ \cite{A_N_Kolmogorov_1968} and for an Iroshnikov-Kraichnan cascade $\delta = 1/2$ \cite{Iroshnikov_1964, Kraichnan:1965zz}. 
The exponent $\eta$ is introduced to accommodate the low-energy behavior of measured CR spectra as in recent works \cite{Maurin:2010zp, Evoli:2015vaa, Derome:2019jfs, Genolini:2019ewc, Boschini:2020jty}, while traditionally set to $\eta=1$. An $\eta<1$ increases the diffusion coefficient at low energies. A negative $\eta$ phenomenologically accounts for physical nonlinear phenomena at low energies, such as the turbulent dissipation and wave damping that produce a very sharp rise of diffusion coefficient at the rigidity less than around 1.5 GV \cite{Ptuskin:2005ax, Zirakashvili:2014zqa}. 
Additionally, two breaks of $\delta$ could be introduced in certain types of models (see below), in which $\delta$ changes to $\delta_l$ and $\delta_h$ when $\rho<\rho_l$ and $\rho>\rho_h$ for each break respectively, with $\rho_l\sim 4\,\mathrm{GV}$ and $\rho_h\sim 300\,\mathrm{GV}$ typically. 

The momentum-space diffusion coefficient $D_{pp}$, describing the effect of diffusive re-acceleration, is parametrized as follows: 
\begin{equation}
    D_{pp} = \frac{4V_a^2 p^2}{3D_{xx} \delta(4-\delta^2)(4-\delta)} \ ,
    \label{eq:Dpp}
\end{equation}
where $V_a$ is the Alfv\'{e}n velocity, which characterizes the propagation of disturbances in GMFs. The scattering of charged particles by the random motion of the magnetic fields characterized by the Alfv\'{e}n velocity leads to a certain amount of second-order Fermi acceleration during propagation, which can significantly modify the low-energy CR spectra. We consider $V_a$ to be constant throughout the diffusion halo, which should be understood as an effective parameter. In some semi-analytic frameworks, the re-acceleration is assumed to be confined in the Galactic disc and normalized to a half-width of $h\sim0.1\,\mathrm{kpc}$, which allows for fast calculations for CR propagation \cite{maurin2002galacticcosmicraynuclei}. The value of $V_a$ obtained in this semi-analytical approach should be roughly rescaled by a factor of $\sqrt{h/z_\mathrm{h}}$ when compared with the one adopted in this work. 

The CR source term $q$ in Eq.~\eqref{eq: prop eq} includes both the primary and secondary components. The primary CRs are believed to be accelerated by supernova remnants (SNRs) and pulsar wind nebulae, whose source term is modeled as the product of a broken power-law rigidity spectrum and a spatial distribution of the primary source $n(\vec{r})$. Assuming the spectrum has $m$ breaks at $\rho_i$ ($i=0,1,2,\dots,m-1$), spectral indices $\gamma_{i}$ and $\gamma_{i+1}$ below and above $\rho_i$, the primary source term is written as
\begin{equation}
    q(\vec{r},\, p) = n(\vec{r}) \ \left(\frac{\rho}{\rho_0}\right)^{-\gamma_0}\ \prod\limits_{i=0}^{m-1} \left[ \frac{\mathrm{max}(\rho, \rho_i)}{\rho_i} \right]^{\gamma_i-\gamma_{i+1}} \ ,
    \label{eq: CR source}
\end{equation}
where the spatial distribution $n(\vec{r})=n(R,z)$ follows the distribution of SNRs or pulsars \cite{case1996revisiting}, given by 
\begin{equation}
    n(R,z) \propto \left(\frac{R}{R_\odot}\right)^a \exp\left( -b \frac{R}{R_\odot} \right) \exp\left( -\frac{|z|}{z_0} \right) \ ,   \label{eq: CR spatial}
\end{equation}
where $a=1.9$, $b=5.0$ \cite{Lorimer:2006qs}, $z_0=0.2\,\mathrm{kpc}$, and $R_\odot=8.5\,\mathrm{kpc}$. We enforce $n(R,z)$ remained constant from $R=10\,\mathrm{kpc}$ and fixed to 0 for $R>15\,\mathrm{kpc}$. Due to linearity of Eq.~\eqref{eq: prop eq} in $\psi$ (excluding the source term $q$), the normalization of primary source term is absorbed into the proton flux normalization or primary isotopic abundances, which are free parameters of the model. 

Secondary CRs arise from the interaction between primary CRs and interstellar gas during propagation. For a secondary species $s$, the source term is given by 
\begin{equation}
    q_s(\vec{r}, p) = \sum_j n_j(\vec{r}) \sum_i \int \mathrm{d}p_i\, c\beta_i \psi_i(\vec{r}, p_i) \frac{\mathrm{d}\sigma_{ij\rightarrow s}}{\mathrm{d}p}(p, p_i) \ ,
    \label{eq:secondary_source}
\end{equation}
where $i$ indexes primary CR species producing the species $s$, $j$ indexes interstellar gas components, $n_j$ is number density of the gas, and $\mathrm{d}\sigma_{ij\rightarrow s}/\mathrm{d}p$ is the differential production cross section. The interstellar gas consists mostly of hydrogen (H) and helium (He) with a number ratio of 9:1. H has different states: atomic (H\,I), molecular ($\mathrm{H}_2$) and ionized (H\,II), and He is mostly neutral. In practical \texttt{Galprop} calculations, $j$ only runs over H\,I, $\mathrm{H}_2$, and H\,II, while the contributions from interactions with He are accounted for by rescaling the corresponding H interaction cross sections with effective factors. 

PBHs in the Galaxy can be regarded as stable sources of CR all-electrons due to Hawking radiation. 
Assuming PBHs constitute a fraction $f_\mathrm{PBH}$ of the DM, their density profile follows that of the DM as $\rho_\mathrm{PBH}(\vec{r}) = f_\mathrm{PBH} \rho_\mathrm{DM}(\vec{r})$. We adopt NFW DM profile \cite{Navarro:1996gj} with scale radius $r_s=24.42\,\mathrm{kpc}$, and Burkert DM profile \cite{Burkert:1995yz} with scale radius $r_s=12.67\,\mathrm{kpc}$. Both DM profiles are normalized such that $\rho_\mathrm{DM}(r_\odot) = 0.4\,\mathrm{GeV/cm}^3$ at the Solar position \cite{deSalas:2020hbh}. The NFW profile is cuspy towards the Galactic center, while the Burkert profile exhibits a constant density core, and predicts a smaller DM density in the innermost part of the Galaxy. The source term for the evaporated all-electrons, with a mass function $\mathrm{d}n/\mathrm{d}M_\mathrm{PBH}$ of PBHs, is given by 
\begin{equation}
    q_{e^\pm}(\vec{r}, p) = \int_{M_\mathrm{min}}^\infty \frac{\mathrm{d}n}{\mathrm{d}M_\mathrm{PBH}} \beta \frac{\mathrm{d}^2 N_{e^\pm}}{\mathrm{d}t\mathrm{d}E} \, \mathrm{d}M_\mathrm{PBH} \ ,
    \label{eq:pbh_source}
\end{equation}
where $M_\mathrm{min}=5\times 10^{14}\,\mathrm{g}$, $\mathrm{d}^2 N_{e^\pm}/\mathrm{d}t\mathrm{d}E$ is the emission rate of a single PBH of mass $M_\mathrm{PBH}$ (see section~\ref{sec: PBH evaporation}), and $\mathrm{d}n/\mathrm{d}M_\mathrm{PBH}$ is normalized to $\rho_\mathrm{PBH}(\vec{r}) = f_\mathrm{PBH}\,\rho_\mathrm{DM}(\vec{r})$ through Eq.~\eqref{eq:PBH norm}, thus depends on $\vec{r}$. 

\subsection{Cosmic Ray Propagation in the Heliosphere}   \label{subsec: solar modulation}
As CRs propagate from the heliopause (HP), i.e., heliosphere boundary, to the Earth, they undergo affects by heliospheric magnetic fields generated by the solar wind. These physical processes suppress CR spectra below $\sim 50$ GV measured at the Earth compared to the interstellar spectra and introduce time-dependent variations related to solar activities \cite{Potgieter_2013}. This effect is known as solar modulation. Since all CR measurements -- except Voyager-1 (since August 2012) and Voyager-2 (since December 2018) -- are conducted within the heliosphere, CR propagation studies require the modeling of both galactic propagation and solar modulation. The propagation of CRs in the heliosphere is described by the Parker equation \cite{Parker:1965ejd}: 
\begin{equation}
    \frac{\partial U}{\partial t} = \nabla \cdot \left(K^S \nabla U - (\vec{V}_\mathrm{sw} + \vec{v}_\mathrm{d}) U \right) + \frac{1}{3} \nabla \cdot \vec{V}_\mathrm{sw} \frac{\partial}{\partial T} \left( \alpha_\mathrm{rel} T U \right) \ ,    \label{eq: Parker}
\end{equation}
where $U(\vec{r}, t, T)$ is the number density of CR particles per unit kinetic energy $T$, $K^S$ is the symmetric part of diffusion tensor, $\vec{V}_\mathrm{sw}$ is the solar wind velocity, $\vec{v}_\mathrm{d}$ is the magnetic drift velocity, and $\alpha_\mathrm{rel} = (T + 2T_0)/(T + T_0)$ with $T_0$ denoting the rest energy of the CR particle. The public \texttt{Helmod} code \cite{Bobik:2011ig, bobik2016forward, Boschini:2017gic, Boschini:2018zdv, Boschini:2019ubh, Boschini:2022jwz} offers modulation tables based on numerical solutions of Eq.~\eqref{eq: Parker} with parameters calibrated using the up-to-date CR data. These tables relate local interstellar (LIS) and top-of-atmosphere (TOA) CR spectra, denoted by $\Phi^\mathrm{LIS}$ and $\Phi^\mathrm{TOA}$, via 
\begin{equation}
    \Phi^\mathrm{TOA}(\rho) = \int \Phi^\mathrm{LIS}(\rho^\prime) \, G(\rho, \rho^\prime; t)  \, \mathrm{d}\rho^\prime \ ,
\end{equation}
where $G(\rho, \rho^\prime; t)$ describes the probability for a CR particle with initial rigidity $\rho^\prime$ at the heliopause to be observed at TOA with rigidity $\rho$, and $t$ denotes the time interval of the CR measurement. The \texttt{Helmod} code is able to quantitatively reproduce the time variation of CR fluxes, such as that of protons, and the predicted LIS proton flux is in remarkable agreement with the Voyager-1 data \cite{Boschini:2019ubh}. Other numerical codes solving the Parker equation include \texttt{SOLARPROP} \cite{Kappl:2015hxv} and \texttt{HelioProp} \cite{Vittino:2017fuh}, etc. 

Numerically solving the Parker equation is computationally demanding for CR propagation studies. Consequently, the much simpler force-field approximation \cite{Gleeson:1968zza} is widely adopted. This model provides an analytical solution of the Parker equation under a series of assumptions: steady-state system, spherical symmetry, and zero streaming. Assuming the diffusion coefficient is separable $k(r,\rho) = \beta k_1(r) k_2(\rho)$, where $\beta = v/c$ and $r$ is the distance from the Sun, the solution relates LIS and TOA CR spectra via 
\begin{equation}
    \frac{\Phi^\mathrm{TOA}}{(\rho^\mathrm{TOA})^2} = \frac{\Phi^\mathrm{LIS}}{(\rho^\mathrm{LIS})^2} \quad \Longrightarrow \quad \Phi^\mathrm{TOA}(T^\mathrm{TOA}) = \frac{T^\mathrm{TOA}(T^\mathrm{TOA} + 2M)}{T^\mathrm{LIS}(T^\mathrm{LIS} + 2M)} \Phi^\mathrm{LIS}(T^\mathrm{LIS}) \ ,
    \label{eq: force field}
\end{equation}
where $\rho$ is the rigidity, $T$ is the kinetic energy, and $M$ is the particle mass. The rigidities $\rho^\mathrm{LIS}$ and $\rho^\mathrm{TOA}$ are connected through
\begin{equation}
    \int_{\rho^\mathrm{TOA}}^{\rho^\mathrm{LIS}} \frac{\beta k_2(\rho)}{\rho}  \mathrm{d}\rho = \int_{r_\mathrm{TOA}}^{r_\mathrm{HP}} \frac{V_\mathrm{sw}(r)}{3k_1(r)}  \mathrm{d}r \equiv \phi \ ,
\end{equation}
where $\phi$ is the modulation potential. For $k_2(\rho) \propto \rho$, this gives 
\begin{equation}
    T^\mathrm{LIS} = T^\mathrm{TOA} + |Ze| \phi \ ,
\end{equation}
where $|Ze|$ is the absolute charge of the CR particle. In practice, $\phi$ is considered as a free nuisance parameter in the fitting procedure of CR propagation models. 
There exists a strong degeneracy between $\phi$ and other model parameters. 
The validity of force-field approximation is challenged by variation of CR monthly fluxes measured by PAMELA and AMS-02 \cite{Corti:2019jlt}. There are proposed modifications to the force-field approximation by introducing rigidity-dependent \cite{Cholis:2015gna, 2016shin.confE.122C, Gieseler:2017xry} or charge-asymmetric \cite{Kuhlen:2019hqb, Kuhlen:2020zqk} modulation potential. 
However, for CR propagation studies concerning CR data measured over a long time period, the force-field approximation is widely adopted. 

\subsection{Benchmark Propagation Models}  \label{sec: benchmark CR}
In sections~\ref{subsec: CR propagation} and \ref{subsec: solar modulation}, we have introduced the basic physical concepts of CR propagation in the Galaxy and heliosphere. However, the specific parameter values, including those in the diffusion equation Eq.~\eqref{eq: prop eq}, primary CR source term Eq.~\eqref{eq: CR source}, and those for solar modulation (in the parker equation Eq.~\eqref{eq: Parker} or the modulation potential $\phi$), need to be determined through fitting to CR measurement data \cite{Strong:2007nh}. 
Secondary-to-primary nuclei flux ratios, such as the most widely used B/C flux ratio, can probe the grammage (representing the integrated gas density along the CR trajectory before escape from the diffusion halo) of CRs and are primarily sensitive to propagation parameters, as the source dependence largely cancel out. 
It is known that re-acceleration can provide a natural mechanism to reproduce the low-energy B/C flux ratio with Kolmogorov type of turbulence and a spectral break in the power-law spectra of primary CRs at a rigidity of a few GV \cite{Simon:1996dk}. Diffusive re-acceleration models with a significant Alfv\'{e}n velocity $V_a\sim\mathcal{O}(10)\,\mathrm{km/s}$ have received strong support from a number of independent analyses \cite{Trotta:2010mx, Jin:2014ica, Johannesson:2016rlh, Boschini:2017fxq, Boschini:2018baj, Boschini:2020jty, DeLaTorreLuque:2021yfq, Luque:2021nxb}, in which the values of $V_a$ are within 30--45 km/s. Several recent analyses obtained slightly smaller $V_a\sim 20\,\mathrm{km/s}$ \cite{Yuan:2017ozr, Korsmeier:2021brc, Luque:2021nxb, Silver:2024ero}, when including additional model free parameters, such as introducing a low-energy $\delta$ break or a free parameter $\eta$, and considering the high-energy break of $\delta$ around 300 GV. 
Notably, analyses fitting to proton, antiproton, and He data preferred a substantially lower $V_a$ \cite{Johannesson:2016rlh, Korsmeier:2016kha}, suggesting different CR species may sample distinct interstellar environments due to spatially dependent propagation effects. 

Since the effects of propagation parameters are partially degenerated, it is possible to construct alternative CR propagation models without the re-acceleration (i.e., with $V_a$ fixed to zero) but including a low-energy break of $\delta$ at $\rho_l \sim 4\,\mathrm{GV}$. 
These types of diffusion break models can reproduce the similar structure in B/C flux ratio. 
So far, both scenarios can well explain the data of secondary-to-primary nuclei flux ratio \cite{Cummings:2016pdr, Korsmeier:2021brc, Genolini:2019ewc, Weinrich:2020cmw}. 
Notably, it is possible to distinguish the two scenarios by considering additional observables. For example, the existence of significant re-acceleration can be probed by synchrotron and soft $\gamma$-ray emissions \cite{Orlando:2017mvd}, and a break in the injection primary source can be examined by $\gamma$-ray emissions from gas clouds \cite{Neronov:2011wi}. 

For constraining $f_\mathrm{PBH}$, the diffusion halo half-height $z_\mathrm{h}$ is a critical parameter. Since CRs escape freely at the boundary of diffusion halo, only the PBHs within the diffusion halo can contribute to observable signals. Thus, the uncertainty of $z_\mathrm{h}$ can directly affect the constraints on $f_\mathrm{PBH}$. 
Because of the famous degeneracy between $z_\mathrm{h}$ and $D_0$, it is hard to constrain the value of $z_\mathrm{h}$ through fitting to the data of secondary-to-primary nuclei flux ratios. 
The flux ratio of unstable-to-stable secondary isotopes can provide insights into the residence time of CRs within the diffusion halo. The life time of $^{10}\mathrm{Be}$ is about 2 million years, comparable to the typical CR residence time. Hence the $^{10}\mathrm{Be}/^{9}\mathrm{Be}$ flux ratio is an ideal observable for constraining $z_\mathrm{h}$. However, due to the large uncertainty of Be isotopic data (such as the data of ACE-CRIS, ACE-SIS, ISOMAX, and PAMELA) and their relatively low energy upper limit $\lesssim 2\,\mathrm{GeV/n}$, it is still hard to obtain stringent constraints on $z_\mathrm{h}$. As a substitution, the data of Be/B flux ratio were used to constrain $z_\mathrm{h}$ in several recent works. 
For example, in \cite{Evoli:2019iih} a lower bound $z_\mathrm{h}\geq 5\,\mathrm{kpc}$ was set, and in \cite{Weinrich:2020ftb} $z_\mathrm{h}=5^{+3}_{-2} \,\mathrm{kpc}$ was found. A relatively smaller $z_\mathrm{h}=3.8^{+2.8}_{-1.6}\,\mathrm{kpc}$ was found in the AMS-02 Be/B analysis in \cite{Maurin:2022gfm}, and $z_\mathrm{h}=4.7\pm 2\,\mathrm{kpc}$ was obtained using the $^{10}\mathrm{Be}/^{9}\mathrm{Be}$ data in the same work. In \cite{Boschini:2020jty}, $z_\mathrm{h}=4.0\pm0.6\,\mathrm{kpc}$ was obtained using the CR data of light nuclei (with charge number $Z\leq 28$), but the uncertainties of production cross sections of secondary CRs were not considered. 
In 2022 ICHEP, AMS-02 collaboration preliminarily reported the Be isotopic measurements \footnote{\url{https://agenda.infn.it/event/28874/contributions/170166/}}, which have reached an unprecedented energy of 12 GeV/n. However, the uncertainties of the production cross sections of secondary CRs still prevent the community to obtain a robust constraint on $z_\mathrm{h}$. Using the preliminary data of $^{7}\mathrm{Be}$ and $^{10}\mathrm{Be}$ nuclei, $z_\mathrm{h}=5.67\pm 0.76\,\mathrm{kpc}$ was obtained in \cite{Zhao:2024qbj}, where the cross section uncertainties were considered by adopting a cross section parametrization that fully utilizes the available experimental data. 
Besides the uncertainties of production cross sections, it is also reported that the Galactic gas model \cite{DeLaTorreLuque:2024wtv} and inhomogeneous diffusion \cite{Jacobs:2023zch} can affect the constraints on $z_\mathrm{h}$ derived from the $^{10}\mathrm{Be}/^{9}\mathrm{Be}$ data. Given the huge difficulties in constraining $z_\mathrm{h}$, instead of choosing one typical $z_\mathrm{h}$ from the earlier works, we fit parameters of 4 diffusive re-acceleration models with $z_\mathrm{h}$ fixed to 4, 6, 8, 10 kpc respectively, covering $z_\mathrm{h}$ values reported in the literature. With these models, we can illustrate the affect of $z_\mathrm{h}$ on the obtained constraints on $f_\mathrm{PBH}$. 

To discuss the impact of diffusive re-acceleration on the constraints derived from Galactic synchrotron observations, we fit parameters of a set of benchmark CR propagation models to the AMS-02 \cite{AMS:2023anq} and Voyager-1 \cite{Cummings:2016pdr} data on the B/C flux ratio. Besides, in calculations of constraints on $f_\mathrm{PBH}$, we also employ an additional diffusive re-acceleration model with the parameters fitted in \cite{Boschini:2020jty}, in which solar modulation is modeled using the \texttt{Helmod} code \cite{Bobik:2011ig, bobik2016forward, Boschini:2017gic, Boschini:2018zdv, Boschini:2019ubh, Boschini:2022jwz}, to complement our simple treatment of force-field approximation. Our benchmark models are listed as follows: 
\begin{itemize}
    \item \textbf{DRz4-10 models}: four diffusive re-acceleration models with different $z_\mathrm{h}$ (fixed to 4, 6, 8, 10 kpc, respectively). As discussed above, the values of $z_\mathrm{h}$ are selected to illustrate the affect of $z_\mathrm{h}$ on the obtained constraints on $f_\mathrm{PBH}$. We have four free parameters for CR propagation, which are $D_0$ (normalized at $4\ \mathrm{GV}$), $\eta$, $\delta$, and $V_a$. 
    \item \textbf{DBz4 model}: a diffusion break model with $z_\mathrm{h}$ fixed to 4 kpc. We have four free parameters for CR propagation, which are $D_0$ (normalized at $4\ \mathrm{GV}$), $\rho_l$, $\delta_l$ and $\delta$. In this model, $V_a$ is fixed to zero. 
    \item \textbf{GH model}: the model was built with analysis framework of \texttt{Galprop + Helmod} (GH), where the two numerical codes \texttt{Galprop} and \texttt{Helmod} are combined together to provide a single framework to calculate CR fluxes at different modulation levels and at both polarities of the solar magnetic field. This is achieved by an iterative optimization procedure to tune the parameters in both \texttt{Galprop} and \texttt{Helmod} to best reproduce the dataset of CR proton flux measured by PAMELA, BESS, and AMS-02. We adopt the parameters determined in \cite{Boschini:2020jty} which are obtained by a Markov chain Monte Carlo scan of the parameter space to fit the AMS-02 data of light nuclei. 
\end{itemize}

Other parameters of the DRz4-10 and DBz4 models are described below. 
The high-energy spectral hardening around 300 GV has been discovered by several CR measurements \cite{PAMELA:2011mvy, AMS:2018tbl, CALET:2019bmh, DAMPE:2022jgy}. We model this feature by including the high-energy $\delta$ break in the diffusion coefficient $D_{xx}$ \cite{Vladimirov:2011rn, Genolini:2017dfb}, which is described by
\begin{equation}
    D_{xx} = \left\{\begin{aligned} & D_0 \beta^\eta (\rho/\rho_0)^\delta &\text{for } \rho < \rho_h, \\ & D_0 \beta^\eta (\rho_h/\rho_0)^\delta (\rho/\rho_h)^{\delta_h}  &\text{for } \rho > \rho_h, \end{aligned} \right.     \label{eq: high delta break}
\end{equation}
where $\rho_h=308\,\mathrm{GV}$ and $\delta_h=\delta-0.175$ are fixed to the best-fit parameters obtained by \cite{Silver:2024ero}. Because we mostly focus on energies below 10 GeV, and to reduce parameter space dimensionality, we adopt these values from \cite{Silver:2024ero} rather than fit them independently. The low-energy $\delta$ break for the diffusion break model is omitted in Eq.~\eqref{eq: high delta break} for a clear expression. For primary source parameters, we assume one spectral break in the primary CR source term Eq.~\eqref{eq: CR source}, having four free parameters: the break rigidity $\rho_0$, the spectral indices $\gamma_0$ and $\gamma_1$ below and above $\rho_0$, and the primary abundance of $^{12}\mathrm{C}$ isotope $A_\mathrm{C12}$ (with the proton abundance fixed at $1.06 \times 10^6$). For solar modulation, we use the force-field approximation with a single free parameter: modulation potential $\phi$ for the data measured by AMS-02. Altogether we have nine free parameters for both types of models. 

Below, we describe our model fitting procedure, display the estimated model parameters, and show the B/C flux radio predictions for the benchmark models. The numerical details and other fit results are put in appendix \ref{app: CR}. 
In the last part of this section, we show energy spectra of evaporated all-electrons for CR propagation models adopted in this work. 

We adopt Bayesian inference framework to estimate parameter values, with uniform prior probability distribution functions (PDFs) over prior ranges, and a likelihood function in Gaussian form: 
\begin{equation}
    \mathcal{L}(\Theta) = \prod_{d,i} \frac{1}{\sqrt{2\pi\sigma_{d,i}^2}} \exp\left(-\frac{\left(\Phi_d(E_i, \Theta) - \Phi_{d,i}\right)^2}{2\sigma_{d,i}^2}\right)\ ,
\end{equation}
where $\Theta$ denotes model parameters, $d$ indexes CR measurement datasets, $i$ indexes energy bins, $\sigma_{d,i}$ is the experimental uncertainty, $\Phi_d(E_i, \Theta)$ is the model-predicted flux for dataset $d$ at $i$-th energy bin with parameters $\Theta$, and $\Phi_{d,i}$ is the corresponding measured value. Assuming uncorrelated errors, we compute the experimental uncertainty $\sigma_{d,i}$ by adding the reported statistical and systematic errors in quadrature. The posterior PDF, which is the normalized product of prior PDFs and the likelihood function, is sampled with the \texttt{MultiNest} package \cite{Feroz:2007kg, Feroz:2008xx, Feroz:2013hea}. 

\begin{table}
\centering
\begin{tabular}{lllllll}
\hline
Parameter                                & Prior range  & DBz4           & DRz4           & DRz6           & DRz8           & DRz10          \\  \hline
$D_0$ ($10^{28}\,\mathrm{cm^2\,s^{-1}}$) & [1, 10]      & 4.23 (0.18)    & 4.20 (0.17)    & 5.86 (0.23)    & 7.04 (0.28)    & 7.84 (0.31)    \\
$\rho_l$ (GV)                            & [0.5, 10]    & 5.12 (0.36)    & --             & --             & --             & --             \\
$\delta_l$                               & [-1, 0]      & -0.374 (0.090) & --             & --             & --             & --             \\
$\delta$                                 & [0.2, 0.6]   & 0.490 (0.008)  & 0.440 (0.011)  & 0.437 (0.011)  & 0.436 (0.011)  & 0.435 (0.011)  \\
$\eta$                                   & [-2, 2]      & --             & -0.628 (0.137) & -0.607 (0.135) & -0.595 (0.134) & -0.583 (0.138) \\
$V_a$ ($\mathrm{km\,s^{-1}}$)            & [0, 60]      & --             & 21.1 (1.7)     & 20.7 (1.7)     & 20.1 (1.7)     & 19.5 (1.6)     \\
$\gamma_0$                               & [0.2, 3.2]   & 0.885 (0.122)  & 0.848 (0.116)  & 0.869 (0.111)  & 0.879 (0.113)  & 0.885 (0.111)  \\
$\gamma_1$                               & [1.8, 3.2]   & 2.346 (0.008)  & 2.369 (0.008)  & 2.372 (0.008)  & 2.373 (0.008)  & 2.374 (0.008)  \\
$\rho_0$ (GV)                            & [0.1, 20]    & 1.52 (0.11)    & 1.54 (0.10)    & 1.54 (0.10)    & 1.54 (0.10)    & 1.54 (0.09)    \\
$A_\mathrm{C12}$                         & [1000, 6000] & 3487 (28)      & 3421 (28)      & 3410 (28)      & 3407 (28)      & 3404 (28)      \\
$\phi$ (MV)                              & [0, 1500]    & 470 (27)       & 606 (27)       & 606 (26)       & 605 (26)       & 605 (26)       \\
\hline
\end{tabular}
\caption{Model parameters, and their prior ranges, posterior means and standard deviations (show in brackets) of the benchmark models. }
\label{tab: CR_best}
\end{table}

\begin{figure}[htbp]
    \centering
    \includegraphics[width=\textwidth]{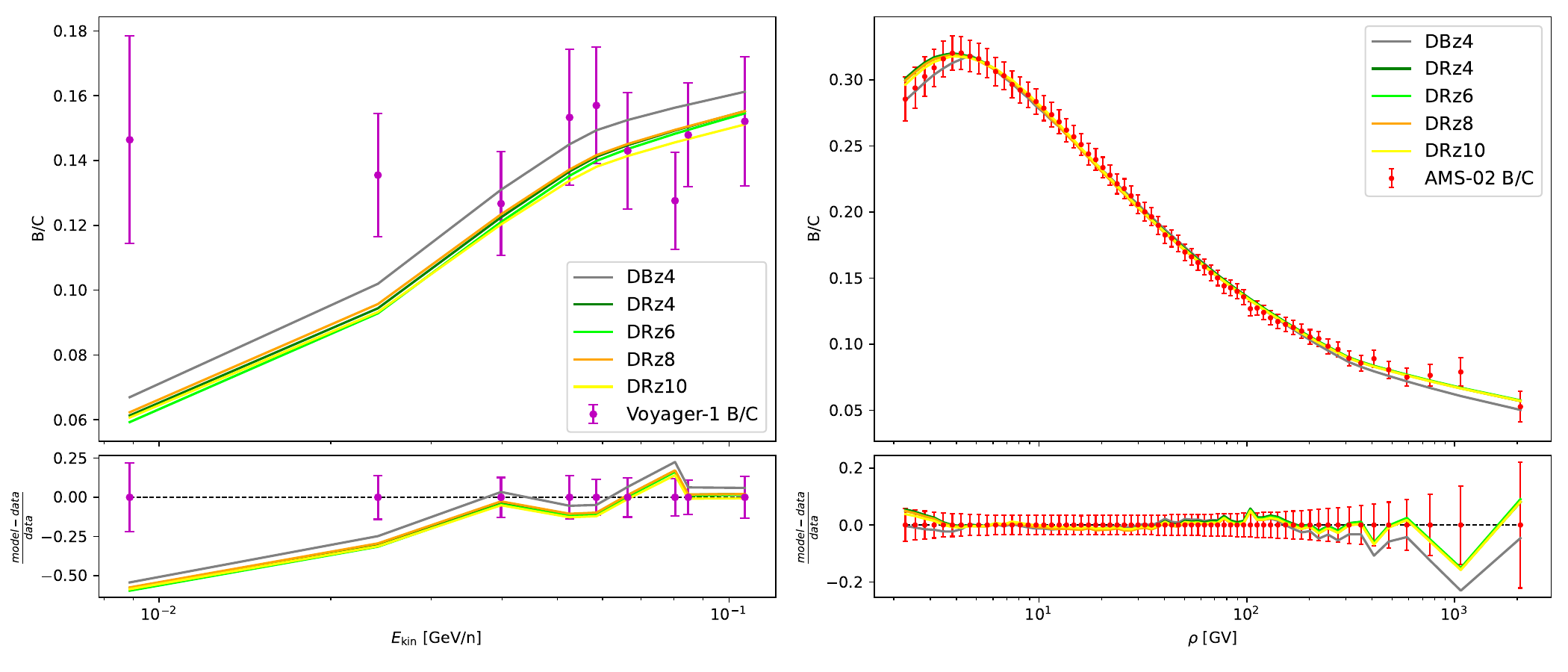}
    \caption{Model predictions and residuals of B/C flux ratio for the benchmark models. Left: for Voyager-1. Right: for AMS-02}
    \label{fig: bestfit_BC}
\end{figure}

In Tab.~\ref{tab: CR_best}, we list the model parameters, and their prior ranges, posterior means and standard deviations of the benchmark models. In Fig.~\ref{fig: bestfit_BC}, we show model predictions and residuals of B/C flux ratio for the benchmark models, and the results for C flux are put in appendix \ref{app: CR}. It can be seen that all DRz4-10 models provide equally good fits to Voyager-1 and AMS-02 data, yielding nearly identical predictions. While the DBz4 model exhibits slightly insufficient spectral hardening around 300 GV, because we fix the $\rho_h$ and $\delta-\delta_h$ same as in diffusive re-acceleration models. Since the value of $\delta$ in diffusion break models is generally larger than that in diffusive re-acceleration models, the fixed parameters for the high-energy $\delta$ break certainly induce some biases in our diffusion break model. A small defect of our models is that they all underpredict the B/C flux ratio data point of Voyager-1 at the lowest energy, a feature also noted in \cite{Cummings:2016pdr, Silver:2024ero}. Nevertheless, our fits show that both types of models can well reproduce the B/C flux ratio data. Notably, our fits obtain significant Alfv\'{e}n velocities $V_a \sim 20\,\mathrm{km/s}$ for all diffusive re-acceleration models. 

The DAMPE experiment has reported high-precision measurements of the B/C flux ratio in the energy range from $10\,\mathrm{GeV/n}$ to $5.6\,\mathrm{TeV/n}$ \cite{DAMPE:2022jgy}, providing stringent constraints on propagation parameters such as $\rho_h$ and $\delta_h$. Because neither our fitting procedure nor that of \cite{Silver:2024ero} incorporated the DAMPE B/C data, we have checked the compatibility of our benchmark models with the DAMPE measurements. 
We find that all models exhibit a mild tendency to underpredict the B/C flux ratio at energies above $\sim 100\,\mathrm{GeV/n}$. For the \mbox{DRz4-10} models, the deviation is at the level of approximately one experimental uncertainty, while for the DBz4 model the maximal deviation reaches roughly twice the experimental uncertainty. Overall, the predicted B/C flux ratios remain broadly consistent with the DAMPE data. 

\begin{figure}[htbp]
    \centering
    \includegraphics[width=\textwidth]{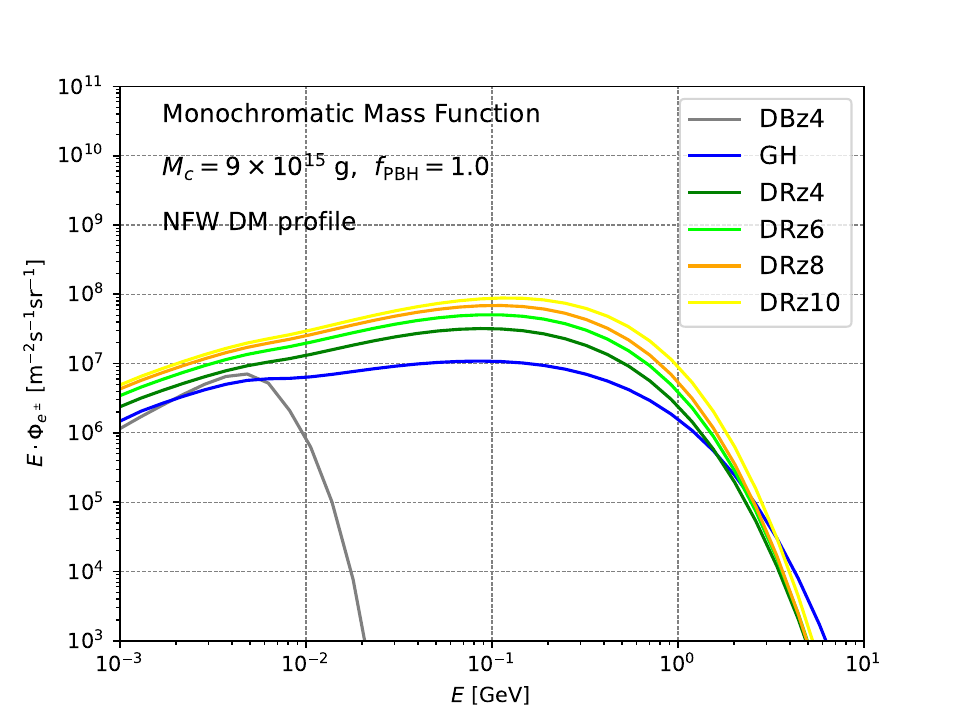}
    \caption{The LIS energy spectra of evaporated all-electrons at the Solar position for CR propagation models adopted in this work, assuming a monochromatic mass function with $M_c=9\times 10^{15}\,\mathrm{g}$, $f_\mathrm{PBH}=1.0$, and NFW DM profile.}
    \label{fig: spec_flux}
\end{figure}

In Tab.~\ref{tab:CR_Params}, we summarize the model parameters relevant to the calculation of synchrotron signals of PBHs, including those obtained from our fits and those of the GH model \cite{Boschini:2020jty}. 
In Fig.~\ref{fig: spec_flux}, we show the LIS energy spectra of evaporated all-electrons at the Solar position for CR propagation models adopted in this work, assuming a monochromatic mass function with $M_c=9\times 10^{15}\,\mathrm{g}$, $f_\mathrm{PBH}=1.0$, and NFW DM profile. 
It can be seen that, for energies above 10 MeV, the all-electron flux of the diffusion break model (DBz4) drops steeply, while those of diffusive re-acceleration models (DRz4-10 and GH) decrease much slower as the energy increasing. 
We show that, for $V_a\sim 20\,\mathrm{km/s}$, a significant fraction of evaporated all-electrons can be boosted to energies of $\sim 100\,\mathrm{MeV}$. 
As $z_\mathrm{h}$ increasing from 4kpc to 10 kpc in DRz4-10 models, the all-electron flux increases by about a factor of 3 at energies of $\sim 100\,\mathrm{MeV}$. 
Notably, compared with the DRz4 model, the GH model with the same $z_\mathrm{h}=4\,\mathrm{kpc}$ predicts a smaller all-electron flux at energies of $\sim 100\,\mathrm{MeV}$, despite having a larger $V_a = 30\,\mathrm{km/s}$. 
While for larger energies of $\sim 1\,\mathrm{GeV}$, the all-electron fluxes of these two models are in agreement. 
This feature may arise from the different values of $\delta$ in these two models, which can affect spectral slope of the LIS all-electron spectra predicted by the models. 
The very distinct values of $\eta$, however, have no effect for these highly relativistic energies of all-electrons. 
The impact of this feature also shows up in the resulting constraints on $f_\mathrm{PBH}$, as will be discussed in section~\ref{sec: fPBH constraints}. 

\begin{table}
\begin{tabular}{lcccccc}
\hline
Parameter                                & DBz4    & GH \cite{Boschini:2020jty}   & DRz4    & DRz6    & DRz8    & DRz10   \\ \hline
$z_\mathrm{h}$ (kpc)                     & 4       & 4     & 4       & 6       & 8       & 10      \\
$D_0$ ($10^{28}\,\mathrm{cm^2\,s^{-1}}$) & 4.23    & 4.3   & 4.20    & 5.86    & 7.04    & 7.84    \\
$\rho_l$ (GV)                            & 5.12    & --    & --      & --      & --      & --      \\
$\delta_l$                               & -0.374  & --    & --      & --      & --      & --      \\
$\delta$                                 & 0.490   & 0.415 & 0.440   & 0.437   & 0.436   & 0.435   \\
$\eta$                                   & --      & 0.7   & -0.628  & -0.607  & -0.595  & -0.583  \\
$V_a$ ($\mathrm{km\,s^{-1}}$)            & --      & 30    & 21.1    & 20.7    & 20.1    & 19.5    \\
$\mathrm{d}V/\mathrm{d}z$ ($\mathrm{km\,s^{-1}\,kpc^{-1}}$) & --        & 9.8       & --        & --        & --        & --        \\
\hline
\end{tabular}
\caption{Model parameters relevant to the calculation of synchrotron signals of PBHs for CR propagation models adopted in this work.}
\label{tab:CR_Params}
\end{table}

\section{Synchrotron Emission and GMF Models}  \label{sec: synch and GMF}
\subsection{Basic Physical Processes}   \label{sec: basic pro}
The Galactic synchrotron emission is generated by CR all-electrons during propagation in the GMF, which is the major component of the Galactic radio emission from $\mathcal{O}(10\,\mathrm{MHz})$ to $\mathcal{O}(10\,\mathrm{GHz})$. With a typical GMF strength of $\mathcal{O}(\mathrm{\mu G})$, the spectral observation of Galactic synchrotron emissions in this frequency range can provide indirect measurements of interstellar CR all-electrons in the energy range from $\mathcal{O}(100\,\mathrm{MeV})$ to $\mathcal{O}(10\,\mathrm{GeV})$. We use the \texttt{Galprop} code to calculate synchrotron emissions \cite{Strong:2011wd, Orlando:2013ysa}. Below, we introduce several physical processes related to the calculation of synchrotron emissions and the GMF modeling. 

Let $\vec{B}_\bot$ denote the projection of magnetic field vector onto the plane perpendicular to the line of sight. The synchrotron emissivity (i.e., power per unit volume per unit frequency) of an isotropic distribution of monoenergetic relativistic all-electrons is partially linearly polarized, which has polarized components parallel and perpendicular to the $\vec{B}_\bot$, denoted by $j_\parallel$ and $j_\bot$, as follows \cite{Rybicki:2004hfl}: 
\begin{equation}
    j_\parallel (\nu) = \frac{\sqrt{3}}{2} \frac{q_e^3}{m_e c^2} B_\bot [F(x)-G(x)] \ ,
\end{equation}
\begin{equation}
    j_\bot (\nu) = \frac{\sqrt{3}}{2} \frac{q_e^3}{m_e c^2} B_\bot [F(x)+G(x)] \ ,
\end{equation}
where $q_e$ is the absolute charge of electron, $m_e$ is the electron mass, $c$ is the speed of light, $x = \nu / \nu_c$ with $\nu_c = 3q_eB_\bot \gamma^2/(4\pi m_e c)$, where $\gamma$ is the Lorentz factor. The functions $F(x)$ and $G(x)$ are defined as 
\begin{align}
    F(x) &= x \int^\infty_x K_{5/3}(x^\prime)  \mathrm{d}x^\prime \ , \\
    G(x) &= x K_{2/3}(x) \ ,
\end{align}
where $K_{5/3}$ and $K_{2/3}$ are the modified Bessel functions of orders $5/3$ and $2/3$, respectively. The total emissivity $j_\mathrm{tot}$ and polarized emissivity $j_\mathrm{pol}$ are given by 
\begin{equation}
    j_\mathrm{tot} (\nu) = j_\bot (\nu) + j_\parallel (\nu) \ ,   \label{jtot}
\end{equation}
\begin{equation}
    j_\mathrm{pol} (\nu) =  j_\bot (\nu) - j_\parallel (\nu) \ .   \label{jpol}
\end{equation}
We adopt the IAU coordinate convention for Stokes parameters \cite{gorski1999healpixprimer}. The intrinsic polarization angle $\chi_0$ is measured from the Galactic north direction to the polarization direction (perpendicular to the $\vec{B}_\bot$), increasing towards the Galactic east, with $\chi_0$ varying in $[0,\ \pi)$. The emissivities for Stokes parameters $Q$ and $U$, denoted by $j_Q$ and $j_U$, are given by 
\begin{equation}
    j_Q = j_\mathrm{pol} \cos(2\chi_0) \ ,\quad j_U = j_\mathrm{pol} \sin(2\chi_0) \ . \label{jQU}
\end{equation}

Radiation due to the acceleration of an electron/positron moving in the Coulomb field of an ion is known as bremsstrahlung or free-free emission. The inverse process, in which radiation is absorbed by an electron/positron moving in the Coulomb field of an ion, is known as free-free absorption. For interstellar ionized gas, the free-free emission becomes significant at frequencies above a few GHz, and the free-free absorption affects the radiation below about 100 MHz. As modeled in \texttt{Galprop} \cite{Orlando:2013ysa, Allen_astro}, the free-free opacity $k_{ff}$ and emissivity $e_{ff}$ at frequency $\nu$ are given by 
\begin{align}
    k_{ff}(\nu, n_\mathrm{TE}, T_\mathrm{TE}) &= 0.0178\, g_{ff}(\nu, T_\mathrm{TE}) \frac{n_\mathrm{TE}^2}{\nu^2 T_\mathrm{TE}^{3/2}} \ , \\
    e_{ff}(\nu, n_\mathrm{TE}, T_\mathrm{TE}) &= 5.444 \times 10^{-39} g_{ff}(\nu, T_\mathrm{TE}) \frac{n_\mathrm{TE}^2}{T_\mathrm{TE}^{1/2}} \ ,
\end{align}
where $g_{ff}(\nu, T_\mathrm{TE}) = 10.6 + 1.9\,\log_{10}(T_\mathrm{TE}) - 1.26\,\log_{10}(\nu)$, $n_\mathrm{TE}$ is the number density of thermal electrons, which equals the number density of ionized hydrogen (H\,II), and $T_\mathrm{TE}$ is the temperature of thermal electrons. Since $k_{ff}$ and $e_{ff}$ both scale with the local $n_\mathrm{TE}^2$, the opacity and emissivity must be multiplied by a clumping factor $C$ to capture the effect of gas inhomogeneities on the squared-density. Following \cite{Orlando:2013ysa}, we adopt $T_\mathrm{TE} = 7000\,\mathrm{K}$ and $C=100$. The optical depth $\tau_\nu$ along the line of sight is given by 
\begin{equation}
    \tau_\nu(s) = \int_0^s C\,  k_{ff}(\nu, n_\mathrm{TE}, T_\mathrm{TE})\, \mathrm{d}r \ ,
\end{equation}
where $s$ is the line-of-sight distance, and $n_\mathrm{TE}$ depends on position $r$ along the line of sight. 

The polarization angle of an electromagnetic wave rotates when passing through a magnetized plasma. This effect is known as Faraday rotation. The rotated angle is proportional to the square of the wavelength $\lambda$. For an initial polarization angle $\chi_0$, the polarization angle $\chi$ after passing through the plasma becomes 
\begin{equation}
    \chi = \mathrm{RM} \cdot \lambda^2 + \chi_0 \ ,
\end{equation}
where the rotation measure (RM) quantifies the linear change rate of $\chi-\chi_0$ with respect to $\lambda^2$, and is determined by the plasma property as 
\begin{equation}
    \mathrm{RM} = \frac{q_e^3}{2\pi m_e^2 c^4} \int_\mathrm{los} \mathrm{d}r\, n_\mathrm{TE}\, B_{\parallel} \ ,   \label{eq: RM}
\end{equation}
where $n_\mathrm{TE}$ is the number density of thermal electrons, $B_{\parallel}$ is the magnetic field component parallel to the line of sight, and the subscript LOS stands for line-of-sight integration. 

Let $n_\mathrm{CR}(\vec{r}, \gamma)$ denote the number density of CR all-electrons with Lorentz factor $\gamma$ at position $\vec{r}$. For a detector of area $\mathrm{d}A$ oriented towards a small solid angle $\mathrm{d}\Omega$, the received radiation power per unit frequency is given by 
\begin{equation}
    \frac{\mathrm{d}W}{\mathrm{d}\nu} = \int_\mathrm{los} \frac{\mathrm{d}A}{4\pi r^2} \left(\int j_\mathrm{tot}(\nu, \gamma)\,n_\mathrm{CR}(\vec{r}, \gamma) \mathrm{d}\gamma\right) e^{-\tau_\nu(r)}\,  r^2  \mathrm{d}\Omega  \mathrm{d}r  \ ,
\end{equation}
where the subscript LOS stands for line-of-sight integration, and $r$ is the distance to the detector. The measured intensity $I$ is given by 
\begin{equation}
    I(\nu) = \frac{\mathrm{d}W}{\mathrm{d}A\mathrm{d}\Omega\mathrm{d}\nu} = \frac{1}{4\pi} \int_\mathrm{los} \left(\int j_\mathrm{tot}(\nu, \gamma)  n_\mathrm{CR}(\vec{r}, \gamma) \mathrm{d}\gamma\right) e^{-\tau_\nu(r)} \mathrm{d}r \ .   \label{eq: synchI}
\end{equation}
The polarized intensity can be calculated analogously by replacing $j_\mathrm{tot}$ with $j_\mathrm{pol}$. For Stokes parameters $Q$ and $U$, the effect of Faraday rotation must be considered. It is helpful to introduce the complex polarized intensity: 
\begin{equation}
    \tilde{P} \equiv P e^{2i\chi} = Q + iU \ ,
\end{equation}
where $P=|\tilde{P}|$ is the polarized intensity, and $\chi$ is the polarization angle. The equation follows directly from the fact that the synchrotron emissivity is partially linearly polarized. The measured $\tilde{P}$ is then obtained by replacing $j_\mathrm{tot}$ with $j_\mathrm{pol} e^{2i\chi}$ in Eq.~\eqref{eq: synchI}, given by
\begin{equation}
    \tilde{P}(\nu) = \frac{1}{4\pi} \int_\mathrm{los} \left(\int j_\mathrm{pol}(\nu, \gamma) e^{2i\chi(r)}  n_\mathrm{CR}(\vec{r}, \gamma) \mathrm{d}\gamma\right) e^{-\tau_\nu(r)} \mathrm{d}r \ ,    \label{eq: synchP}
\end{equation}
where $\chi(r) = \mathrm{RM}\cdot \lambda^2 + \chi_0$ is the observed polarization angle of the emission from a volume element at line-of-sight distance $r$, which depends on position due to variations of both $\mathrm{RM}$ and $\chi_0$ along the line of sight. 

It is useful to provide results under the assumption of a constant spectral index $\alpha$ of CR all-electrons, implying $n_\mathrm{CR}(\gamma) = N_0 \gamma^{-\alpha}$. In this case, the integrals over $\gamma$ in Eq.~\eqref{eq: synchI} and Eq.~\eqref{eq: synchP} can be evaluated analytically: 
\begin{align}
    I(\nu) &= \int_\mathrm{los} J_I\, e^{-\tau_\nu(r)}\, \mathrm{d}r   \ , \\
    \tilde{P}(\nu) &= \int_\mathrm{los} J_P \, e^{2i\chi(r)}\, e^{-\tau_\nu(r)}\, \mathrm{d}r   \ ,
\end{align}
with 
\begin{align}
    J_I &= \frac{\sqrt{3}q_e^3N_0}{4\pi m_e c^2(1+\alpha)} B_\bot^{\frac{1+\alpha}{2}} \left(\frac{2\pi \nu m_e c}{3q_e}\right)^{\frac{1-\alpha}{2}} \Gamma\left(\frac{\alpha}{4}+\frac{19}{12}\right) \Gamma\left(\frac{\alpha}{4}-\frac{1}{12}\right) \ , \\
    J_P &= \frac{\sqrt{3}q_e^3N_0}{16\pi m_e c^2} B_\bot^{\frac{1+\alpha}{2}} \left(\frac{2\pi \nu m_e c}{3q_e}\right)^{\frac{1-\alpha}{2}} \Gamma\left(\frac{\alpha}{4}+\frac{7}{12}\right) \Gamma\left(\frac{\alpha}{4}-\frac{1}{12}\right) \ ,
\end{align}
where $\Gamma$ is the Gamma function. For a spectral index $\alpha = 3$, which is approximately true for the CR all-electron spectrum at energies of $\mathcal{O}(10\,\mathrm{GeV})$, both $J_I$ and $J_P$ scale with $N_0 B_\bot^2 \nu^{-1}$, providing intuitions to the frequency spectrum and the magnetic field dependence of the Galactic synchrotron emissions. 

\subsection{GMF Models}   \label{sec: GMF models}
To calculate synchrotron signals of PBHs, a realistic model of the GMF is required. However, our understanding of the GMF is limited by both our position within the Galactic disc and the challenges in interpreting observables: rotation measure (RM), synchrotron intensity ($I$), and polarized synchrotron intensity ($P$). As described above, these observables are line-of-sight integrated quantities which probe different GMF components: the RM probes the weighted average of GMF component parallel to the line of sight, with the density of thermal electrons as the weight; while the synchrotron emission ($I$ and $P$) probes the weighted average of the squared perpendicular GMF component, with the density of CR electrons as the weight. 
The contribution of CR positrons is generally negligible in construction of GMF models, which mainly concerns synchrotron emission observations at 408 MHz and 23 GHz. However, CR positrons can have significant affects on synchrotron emissions at lower frequencies of $\lesssim 100\,\mathrm{MHz}$ as discussed in \cite{Orlando:2017mvd}. 
There are no direct measurements of the GMF that do not depend on other components of the interstellar medium. 
Furthermore, these observables can be contaminated by other physical processes, such as free-free emission and dust emission. 

The interstellar medium is known to be turbulent, leading to fluctuations in the GMF on scales larger than $\sim 100\,\mathrm{pc}$ \cite{Haverkorn_2008, Haverkorn:2014jka}. 
The modeling of turbulent interstellar medium is another challenge in construction of GMF models \cite{Beck:2003dd}. 
To simultaneously explain the observations of RM, $I$, and $P$, in many works, the GMF is modeled with three effective component fields that contribute differently to these observables (for a recent review, see \cite{Jaffe:2019iuk}). 
These are regular field, striated random field \footnote{We use the name same as in \cite{Jansson:2012pc,Jansson:2012rt} to avoid possible ambiguity, while in \cite{Strong:2011wd, Orlando:2013ysa} this field is refer to as anisotropic random field. }, and isotropic random field. 
The regular field maintains coherence on kpc scale and contributes to all observables: RM via Eq.~\eqref{eq: RM}, $I$ via Eq.~\eqref{eq: synchI}, and $P$ via Eq.~\eqref{eq: synchP}. 
The striated random field exhibits random strength variations but maintains alignment with a specific direction (typically but not necessarily, the direction of the regular field). Only its direction and variance of the strength are relevant. The striated random field contributes to $I$ and $P$ but not to RM, since its line-of-sight projections cancel out. 
The isotropic random field varies randomly in both strength and direction, and only the strength variance is relevant. It contributes solely to $I$ due to the isotopic random direction. 

The synchrotron emissivity of the random fields is modeled in several ways. In the \texttt{Galprop}~\cite{Orlando:2013ysa} framework, the striated random field with a field strength root-mean-square $B^\mathrm{str}$ is modeled as two opposing regular fields with same strength of $B^\mathrm{str}$. Its total (polarized) synchrotron emissivity is $j^\mathrm{str}_\mathrm{tot(pol)}(\nu, B_\bot^\mathrm{str}) = 2 j^\mathrm{reg}_\mathrm{tot(pol)}(\nu, B_\bot^\mathrm{str})$, where $j^\mathrm{reg}_\mathrm{tot(pol)}$ is the emissivity of regular field (see Eq.~\eqref{jtot} and Eq.~\eqref{jpol}) with $B_\bot$ replaced by $B_\bot^\mathrm{str}$, the projection of the magnetic field onto the plane perpendicular to the line of sight. 
For the isotropic random field with a field strength root-mean-square $B^\mathrm{ran}$, the total synchrotron emissivity is modeled as the average emissivity of isotropically distributed regular fields with the same strength of $B^\mathrm{ran}$, approximated as in \cite{Strong:2011wd}: 
\begin{equation}
    j^\mathrm{ran}_\mathrm{tot}(\nu, B^\mathrm{ran}) = \frac{2 \sqrt{3} q_e^3}{m_e c^2} B^\mathrm{ran}\, x^2 \left[K_{4/3}K_{1/3} - \frac{3}{5} x (K_{4/3}^2 - K_{1/3}^2)\right] \ ,
\end{equation}
where $x = \nu / \nu_c$ with $\nu_c = 3q_e B^\mathrm{ran}\,\gamma^2/(2\pi m_e c)$, and $K_{4/3}$, $K_{1/3}$ are modified Bessel functions of variable $x$. On the other hand, in the framework of \cite{Jansson:2012pc, Jansson:2012rt}, a constant spectral index $\alpha = 3$ is assumed for CR electrons, simplifying the synchrotron emissivity to 
\begin{equation}
    J \propto \langle (B^\mathrm{reg}_\bot + B^\mathrm{str}_\bot + B^\mathrm{ran}_\bot)^2 \rangle = B^2 \sin^2\theta + \beta B^2 \sin^2\theta + \frac{2}{3} (B^\mathrm{ran})^2 \ ,
\end{equation}
where $\langle\, \rangle$ means average over the random fields, $\theta$ is the angle between the line of sight and the regular field, and $\beta$ is the striated-to-regular emission ratio defined in the articles. 
However, for the all-electrons evaporated by PBHs, their energy spectra deviate from $\alpha=3$ power laws. 
We match the model parameters of this scenario with those of the \texttt{Galprop} scenario under the assumption of $\alpha=3$, and compare synchrotron spectra predicted by these two model scenarios. 
The two scenarios obtain nearly identical results for CR electron spectrum with $\alpha \approx 3$. 
While for several spectra of evaporated all-electrons after propagation, the maximal deviation reaches $\sim 20\%$ within the frequency range from 1 MHz to 10 GHz. 
The predictions of the two model scenarios are in agreement, so that we can use the GMF model constructed in \cite{Jansson:2012pc, Jansson:2012rt} to calculate the synchrotron signals of PBHs with the \texttt{Galprop} code. 

The turbulent magnetic fields remain nearly coherent over the small scale of $\sim 100\,\mathrm{pc}$, implying that stochastic fluctuations may not fully average out on kpc scale. 
The model scenarios described above, which capture only the mean effect of the turbulent magnetic fields, neglect their stochastic features. 
A more elaborate modeling method involves stochastic field realizations. 
For example, as implemented in the \texttt{Hammurabi} code \cite{Waelkens:2008gp, Wang_2020}, the turbulent magnetic fields are modeled with realizations of Gaussian stochastic fields with the power spectrum informed by magnetohydrodynamic theory. 
In such framework, observational data correspond to the prediction of a possible stochastic field realization. 
Consequently, a precise match of the model prediction to the observational data is not expected. 
However, such models can predict the level of variation induced by the turbulent magnetic fields, referred to as the the galactic variance. 
The galactic variance itself is an observable and can be used to quantify the statistical significance of the model residuals \cite{Planck:2016gdp}. 
Owing to the high computational cost of stochastic field realizations, few GMF models are constructed with this method, notable examples like \cite{jaffe2010, Jaffe:2011qw, Jaffe:2013yi} are restricted to the Galactic plane. 
For our purpose of computing the synchrotron signals of PBHs, the model capturing the mean feature of turbulent magnetic fields suffices. 
Nevertheless, the stochastic nature of turbulent magnetic fields can introduce additional uncertainties. 
For instance, in the sky map with a pixel resolution of $3.7^\circ$ (with $N_\mathrm{side}=16$ in \texttt{HEALPix} scheme), fluctuations on the scale of 100 pc cannot average out within $\sim 1.5$ kpc far from the observer. 
These remained local fluctuations act as foregrounds of observational data, resulting in an unconsidered uncertainty in our analysis. 

Despite the modeling difficulties, several GMF models have been constructed under specific assumptions about interstellar gas, CR electron density, and GMF geometrical structure \cite{Jaffe:2019iuk}. In this work, we adopt two benchmark models which include all three GMF component fields (regular, striated random, and isotropic random) and achieve agreements with RM and synchrotron (both $I$ and $P$) observational data, as follows: 
\begin{itemize}
    \item \textbf{SUNE model}: the model of the same name constructed in \cite{Orlando:2013ysa}. It utilizes the best-fit regular field from \cite{Sun:2007mx, Sun:2010sm}, including the ASS+RING disc field and an updated toroidal halo field from \cite{Sun:2010sm}. Its striated random field aligns with the regular field and maintains a constant strength ratio relative to the strength of regular field through out the Galaxy. The isotropic random field decreases exponentially in both radial and vertical directions. The random field parameters are fitted using the radio continuum surveys from 22 MHz to 2.3 GHz, and the total and polarized intensity data from 7-year Wilkinson Microwave Anisotropy Probe (WMAP) at frequencies between 23 GHz and 94 GHz. The WMAP MCMC templates are adopted to account for dust and spinning-dust emissions. Free-free absorption and emission are computed with specific interstellar ionized gas model, and the CR electron density follows the z04LMPDS model in \cite{Strong:2010pr}. 
    \item \textbf{JF12 model}: a more complex model constructed in \cite{Jansson:2012pc,Jansson:2012rt}. Its regular field includes: 1. a disc field with eight spiral arms, 2. a toroidal halo field, 3. an X-shaped halo field extending perpendicular to the Galactic plane. The striated random field is parameterized by a parameter $\beta$ in the whole galaxy, defined as the striated-to-regular emission ratio. We use the updated value of $\beta$ in \cite{Jansson:2012rt}. The isotropic random field includes: 1. a disc field mirroring the structure of the regular disc field, 2. a halo field combining radial exponential and vertical Gaussian profiles. The random field parameters are fitted using the total and polarized intensity data from the 7-year WMAP at 23 GHz, and with the assumption of a constant spectral index $\alpha=3$ for CR electron density. 
\end{itemize}

\section{Constraints on the PBH Abundance}  \label{sec: fPBH}

\subsection{Observational Data and Error Estimation}   \label{sec: radio obs and err}
The diffuse Galactic radio emissions arise from several astrophysical processes. At frequencies below a few GHz, synchrotron radiation produced by CR electrons propagating in the GMF dominates the sky brightness. 
At higher frequencies, free-free emission from ionized gas and thermal emission from interstellar dust become increasingly important, while anomalous microwave emission from spinning dust grains may also contribute in the GHz range. 
In this work, we focus on the low-frequency regime, where the observed radio emissions are primarily of synchrotron origin. Moreover, low-frequency radio waves can be affected by other astrophysical processes when passing through the interstellar plasma. 
In particular, free-free absorption can attenuate the emission near the Galactic plane at frequencies below $\sim 100\,\mathrm{MHz}$, while Faraday rotation can lead to significant depolarization below $\sim 1\,\mathrm{GHz}$. 

After propagation, the energies of evaporated all-electrons are mostly below a few GeV (see Fig.~\ref{fig: spec_flux}). Thus, the synchrotron signals of PBHs are primarily below a few GHz. We use the data from radio continuum surveys at frequencies between 22 MHz and 1.4 GHz to constrain $f_\mathrm{PBH}$. The datasets include: 
\begin{itemize}
    \item \textbf{22 MHz}: Dominion Radio Astrophysical Observatory Northern Hemisphere survey \cite{Roger:1999jy}; 
    \item \textbf{45 MHz}: Northern sky: Japanese Middle and Upper Atmosphere radar array \cite{north_45MHz}, Southern sky: Maipú Radio Astronomy Observatory 45-MHz array \cite{south_45MHz}, and the combined all-sky map from \cite{Guzman:2010da}; 
    \item \textbf{85 MHz}: Parkes radio telescope \cite{LandW_150MHz}; 
    \item \textbf{150 MHz}: Parkes-Jodrell Bank all-sky survey \cite{LandW_150MHz}; 
    \item \textbf{408 MHz}: Bonn-Jodrell Bank-Parkes all-sky survey \cite{Haslam:1982zz}, and reprocessed by \cite{Remazeilles:2014mba}; 
    \item \textbf{850 MHz}: Dwingeloo telescope \cite{Berk_850MHz}; 
    \item \textbf{1420 MHz}: Northern sky: 25-m Stockert telescope \cite{SVE_1420MHz_N1, SVE_1420MHz_N2}, Southern sky: 30-m Villa-Elisa telescope \cite{SVE_1420MHz_S1, SVE_1420MHz_S2}. 
\end{itemize}
All datasets are publicly accessible on the LAMBDA website \footnote{\url{https://lambda.gsfc.nasa.gov/product/foreground/fg_diffuse.html}}. The observed intensity is recorded in the sky map of brightness temperature $T(\nu)$, which is converted from the intensity $I(\nu)$ as 
\begin{equation}
    T(\nu) = \frac{c^2\,I(\nu)}{2k_\mathrm{B}\,\nu^2} \ ,
\end{equation}
where $c$ is the speed of light, $k_\mathrm{B}$ is the Boltzmann constant, and $\nu$ is the frequency. It is the temperature that a body with a Rayleight-Jeans spectrum would need to emit the same intensity at the frequency $\nu$. Thus, the brightness temperature is defined in unit of Kelvin ($\mathrm{K_{br}}$). To obtain conservative constraints on $f_\mathrm{PBH}$, we do not subtract any offset from the observational sky maps. 

All observational sky maps are pixelated using the \texttt{HEALPix} scheme \cite{gorski1999healpixprimer}, of which the minimum $N_\mathrm{side}=64$. To suppress the effect of fluctuations of the turbulent magnetic fields and to estimate the observational errors, we degrade the sky maps to $N_\mathrm{side}=16$, corresponding to a resolution of approximately $3.7^\circ$. 
The observational error for each degraded pixel is estimated as the standard deviation of values of all high-resolution sub-pixels contained within the degraded pixel. As discussed in \cite{Jansson:2012pc, Jansson:2012rt}, this error estimation method not only accounts for experimental errors, but also includes the astrophysical variance caused by turbulent magnetic fields and inhomogeneities in the interstellar medium. 
For sky maps with the lowest original resolution ($N_\mathrm{side}=64$), the standard deviation is computed with 16 sub-pixels contained within each degraded pixel. For robustness of our constraints, we reject pixels (with $N_\mathrm{side}=16$) which contain sub-pixels with bad-values at the original resolution.

\subsection{Synchrotron Signals of PBHs}   \label{sec: synch prediction}

\begin{figure}[htbp]
    \centering
    \includegraphics[width=\textwidth]{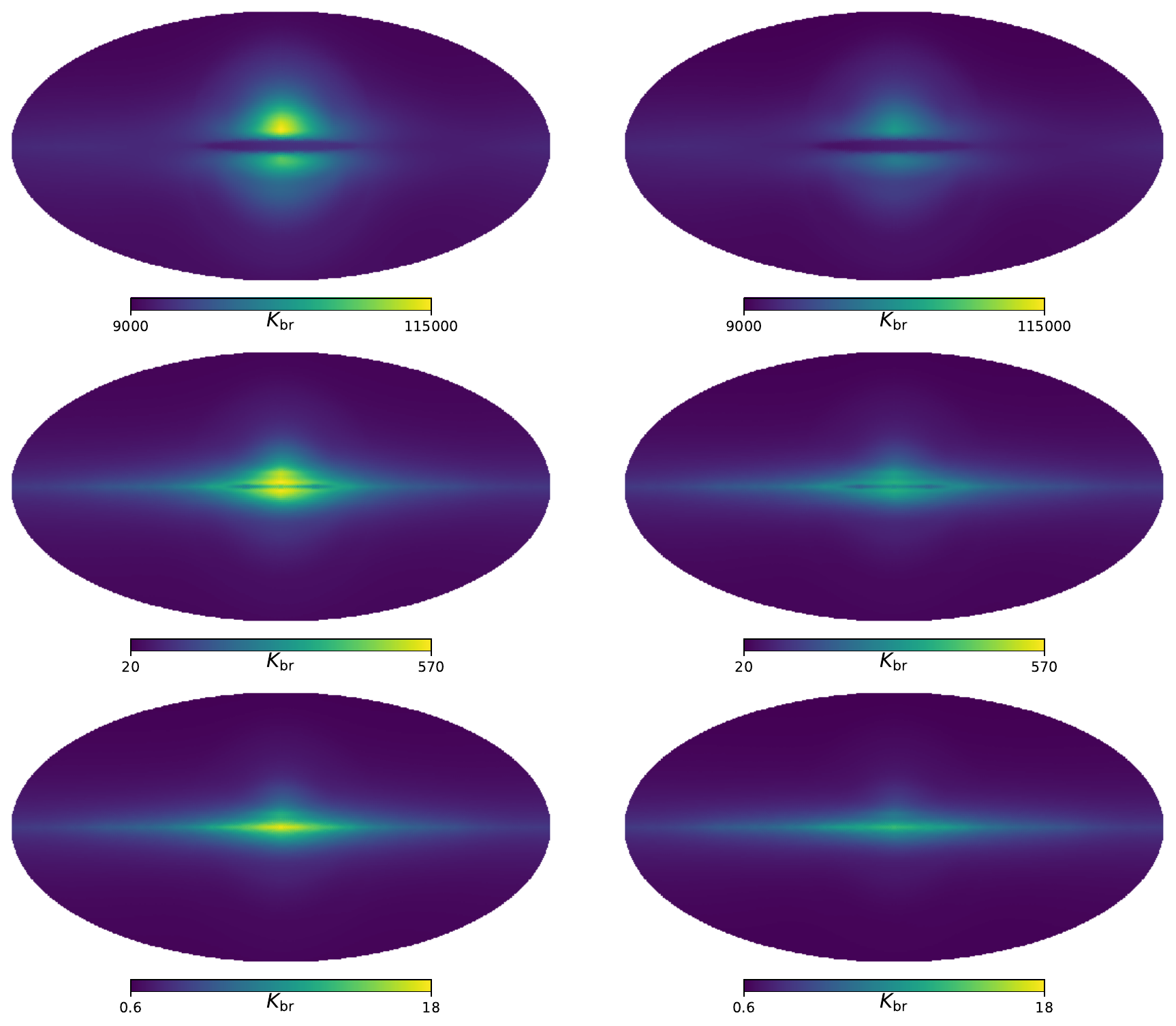}
    \caption{Sky maps of synchrotron signals of PBHs with NFW and Burkert DM profiles (left and right) at 22 MHz, 150 MHz, and 408 MHz (top to bottom), assuming a monochromatic mass function with $M_c=5\times 10^{14}\,\mathrm{g}$, GH CR propagation model, and SUNE GMF model. All maps are computed with $f_\mathrm{PBH}=10^{-8}$, and are plotted with $N_\mathrm{side}=64$ and linear color mapping. }
    \label{fig: sky_DM_SUNE}
\end{figure}

\begin{figure}[htbp]
    \centering
    \includegraphics[width=\textwidth]{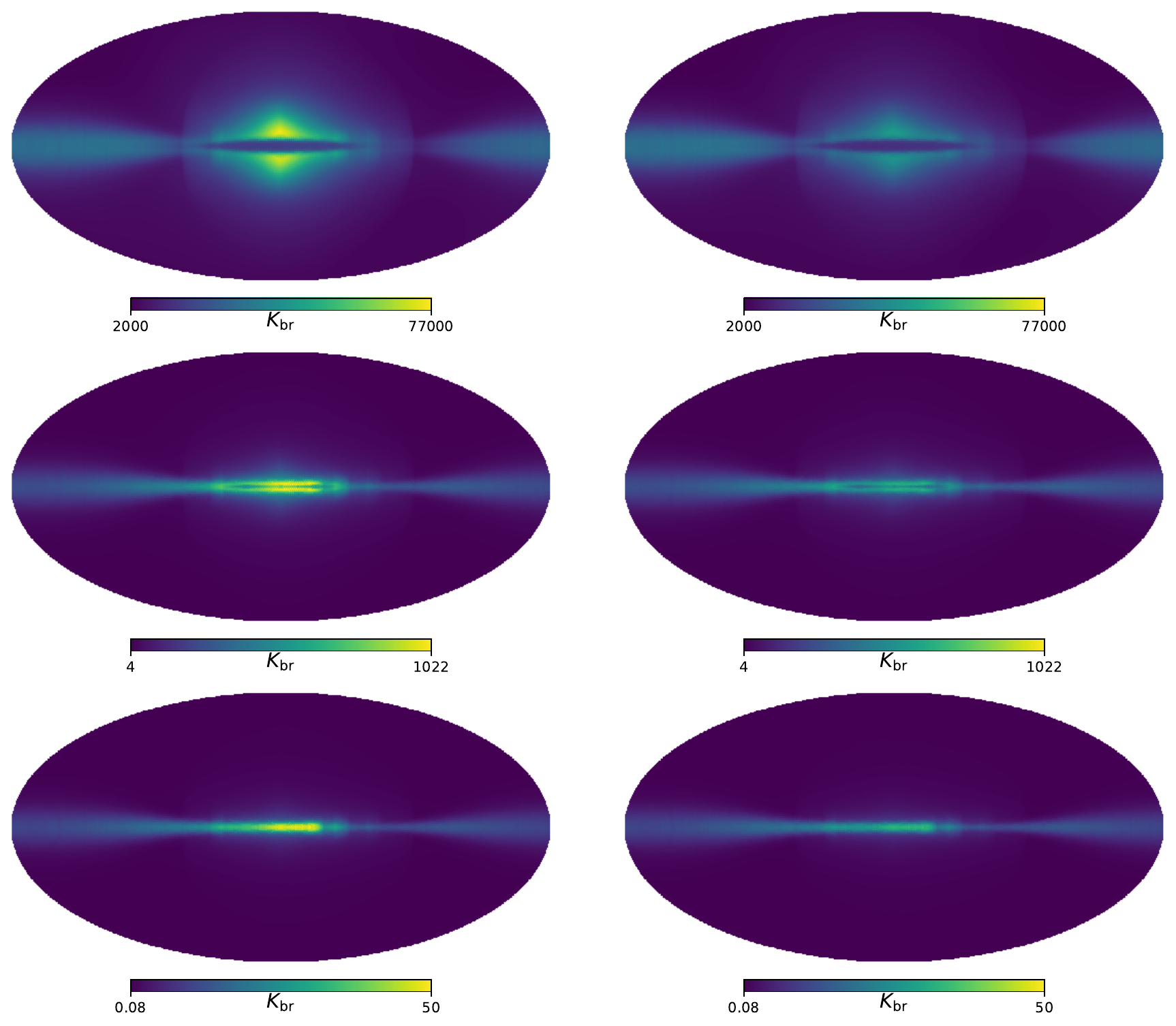}
    \caption{Same as Fig.~\ref{fig: sky_DM_SUNE} but for JF12 GMF model.}
    \label{fig: sky_DM_JF12}
\end{figure}

In calculation of synchrotron signals of PBHs, we can safely neglect the extragalactic contributions to the signals and consider only the synchrotron emissions produced within the Galaxy. This is due to the very small averaged magnetic field strength $B\sim \mathrm{nG}$ on cosmological scales \cite{Amaral:2021mly}, which strongly suppresses the emissivity since it scales approximately as $\propto B^2$. 
In Fig.~\ref{fig: sky_DM_SUNE} and Fig.~\ref{fig: sky_DM_JF12}, we show sky maps of synchrotron signals of PBHs with NFW and Burkert DM profiles (left and right) at 22 MHz, 150 MHz, and 408 MHz (top to bottom), assuming a monochromatic mass function with $M_c=5\times 10^{14}\,\mathrm{g}$, GH CR propagation model, and SUNE and JF12 GMF models, respectively. All maps are computed with $f_\mathrm{PBH}=10^{-8}$. 
The 22 MHz and 150 MHz maps correspond to the frequencies at which the strongest single-frequency constraints on $f_\mathrm{PBH}$ are obtained for the SUNE and JF12 models, respectively (see Fig.~\ref{fig: fPBH-freq} below). 
The 408 MHz map is shown because it corresponds to the frequency at which the observational sky map is minimally contaminated by non-synchrotron processes. 
Dark regions on the Galactic disc towards the Galactic center in the maps at 22 MHz and 150 MHz arise from free-free absorption, which diminishes at higher frequencies and becomes negligible at 408 MHz. 
It can be seen that Burkert profile predicts weaker synchrotron signals than those predicted by NFW profile towards the region near the Galactic center, where the Burkert profile predicts a smaller DM density. 

\begin{figure}[htbp]
    \centering
    \includegraphics[width=\textwidth]{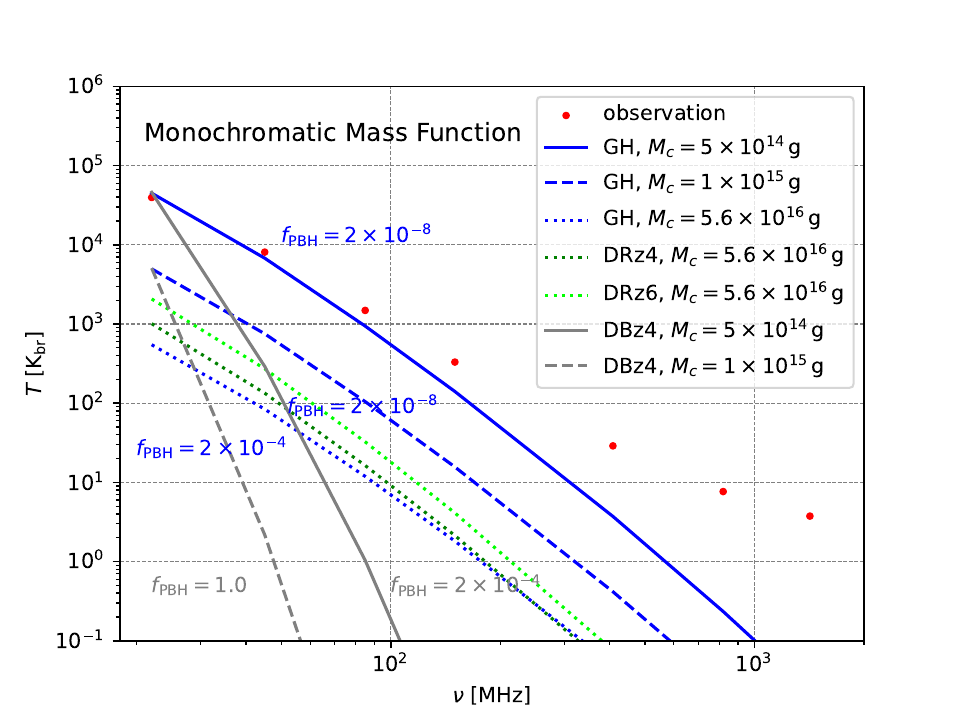}
    \caption{Spectra of synchrotron signals of PBHs for GH, DRz4, DRz6 and DBz4 CR propagation models, assuming SUNE GMF model, NFW DM profile, and monochromatic mass functions of selected $M_c=5\times 10^{14}\,\mathrm{g}$, $1\times10^{15}\,\mathrm{g}$, and $5.6\times 10^{16}\,\mathrm{g}$. The spectra are averaged within the intermediate latitudes $10^\circ<|b|<40^\circ$. Observed intensities averaged within the same region are also shown. For a tight plot, the spectra of GH model are scaled by $f_\mathrm{PBH}=2\times 10^{-8}$, $2\times 10^{-8}$ and $2\times 10^{-4}$ for each $M_c$, and the spectra of DBz4 model are scaled by $f_\mathrm{PBH}=2\times 10^{-4}$ and $1.0$ for the smaller two $M_c$ (as labeled). Only the spectra with $M_c=5.6\times 10^{16}\,\mathrm{g}$ for DRz4 and DRz6 models are shown, which are also scaled by $f_\mathrm{PBH}=2\times 10^{-4}$ as the spectrum of GH model with the same $M_c$ (the $f_\mathrm{PBH}$ is not labeled). }
    \label{fig: synch spec}
\end{figure}

In Fig.~\ref{fig: synch spec}, we show the spectra of synchrotron signals of PBHs for GH, DRz4, DRz6 and DBz4 CR propagation models, assuming SUNE GMF model, NFW DM profile, and monochromatic mass functions of selected $M_c=5\times 10^{14}\,\mathrm{g}$, $1\times10^{15}\,\mathrm{g}$, and $5.6\times 10^{16}\,\mathrm{g}$. The spectra are averaged within the intermediate latitudes $10^\circ<|b|<40^\circ$. Observed intensities averaged within the same region are also shown. The spectrum of DBz4 model with $M_c=5.6\times 10^{16}\,\mathrm{g}$ is much lower than the observed intensities, thus is not shown. For a tight plot, the spectra of GH model are scaled by $f_\mathrm{PBH}=2\times 10^{-8}$, $2\times 10^{-8}$ and $2\times 10^{-4}$ for each $M_c$, and the spectra of DBz4 model are scaled by $f_\mathrm{PBH}=2\times 10^{-4}$ and $1.0$ for the smaller two $M_c$ (as labeled in the figure). 
It can be seen that the synchrotron signal intensity of GH model is larger than that of DBz4 model significantly, by a factor of $10^4$ for $M_c=5\times 10^{14}\,\mathrm{g}$ at the lowest observational frequency 22 MHz, and the factor becomes much larger for higher frequencies and larger $M_c$. 
By requiring the synchrotron signals not exceeding the observed intensities, constraints on $f_\mathrm{PBH}$ can be roughly estimated. Since the spectra of synchrotron signals are softer than the observed spectrum, we expect that the observational data at lower frequencies can set stronger constraints. 
For the GH model, strong constraints can be estimated, such as $f_\mathrm{PBH}<2\times 10^{-8}$ for $M_c=5\times 10^{14}\,\mathrm{g}$, and one order of magnitude weaker constraint for $M_c=1\times 10^{15}\,\mathrm{g}$. Even for heavy PBHs with $M_c=5.6\times 10^{16}\,\mathrm{g}$, $f_\mathrm{PBH}<2\times 10^{-2}$ can be estimated. 
In contrast, for the DBz4 model, only the lightest PBHs with $M_c < 1\times 10^{15}\,\mathrm{g}$ can be constrained. The estimated constraints are much weaker, for example, $f_\mathrm{PBH}<2\times 10^{-4}$ for $M_c=5\times 10^{14}\,\mathrm{g}$. 

In Fig.~\ref{fig: synch spec}, we also show the spectra of synchrotron signals of PBHs for DRz4 and DRz6 models with $M_c=5.6\times 10^{16}\,\mathrm{g}$, which are scaled by $f_\mathrm{PBH}=2\times 10^{-4}$ (not labeled in the figure). 
Compared to the GH and DRz4 models whose diffusion halo half-height $z_\mathrm{h}=4\,\mathrm{kpc}$, the DRz6 model predicts larger synchrotron signal intensities for $M_c=5.6\times 10^{16}\,\mathrm{g}$ due to its larger $z_\mathrm{h}=6\,\mathrm{kpc}$. 
The synchrotron signal intensity of the DRz4 model is larger that of the GH model for the same $M_c$ at frequencies of $\mathcal{O}(10\,\mathrm{MHz})$, which arise from the larger flux of evaporated all-electrons at energies of $\mathcal{O}(100\,\mathrm{MeV})$ predicted by the DRz4 model (see Fig.~\ref{fig: spec_flux}). 
For higher frequencies of $\gtrsim 100\,\mathrm{MHz}$, the DRz4 and GH models predict nearly equal synchrotron signal intensities, because the fluxes of evaporated all-electrons predicted by these two models become in agreement at energies of $\gtrsim 1\,\mathrm{GeV}$. 

At this point, we would like to clarify that the inclusion of PBHs as additional CR sources does not affect the self-consistency of the adopted GMF models, which are generally derived under the assumption of a specific CR electron spectrum. 
The parameters of the GMF model are tightly constrained by RMs and high-frequency radio/microwave observations (e.g., $\gtrsim 20\,\mathrm{GHz}$ for WMAP). 
Among these observables, the RMs are completely independent of CR models, and the high-frequency radio emissions are dominated by CR electrons with energies above $\sim 10\,\mathrm{GeV}$. 
As we show in Fig.~\ref{fig: spec_flux}, for the all-electrons evaporated by PBHs, their energy cannot reach 10 GeV even though the re-acceleration effect is considered in propagation. Therefore, the inclusion of PBHs as additional CR sources affects the high-energy CR electron spectrum marginally in the high-energy regime relevant for the calibration of GMF model parameters. We also have checked its correctness by extending the synchrotron signal spectra to higher frequencies of $\sim 20\,\mathrm{GHz}$, and found that the signals drop rapidly above a few GHz and become much smaller than the observed radio emissions.

\subsection{Results}   \label{sec: fPBH constraints}
We constrain $f_\mathrm{PBH}$ by requiring that, for each pixel in the observational sky maps (with $N_\mathrm{side}=16$), the synchrotron signal of PBHs does not exceed the observed intensity plus twice the observational error. Since the synchrotron signals are proportional to $f_\mathrm{PBH}$, we rescale $f_\mathrm{PBH}$ until the limit condition is satisfied. 
To obtain conservative constraints, we do not attempt to model or subtract any astrophysical background components. Instead, we conservatively allow the entire observed radio intensity to be accounted for by synchrotron signals of PBHs. 

\begin{figure}[htbp]
    \centering
    \includegraphics[width=\textwidth]{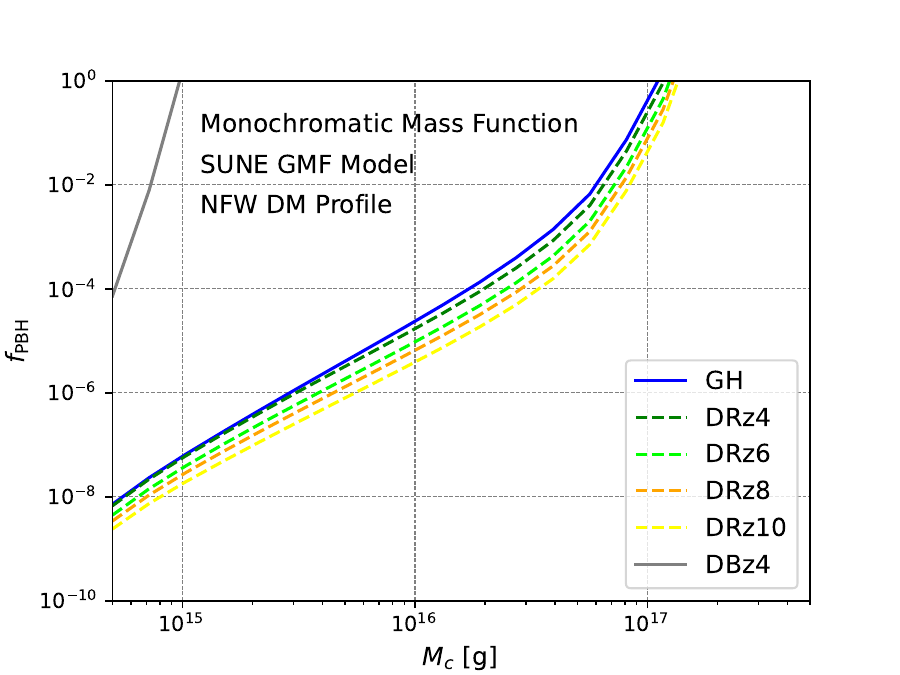}
    \caption{Constraints on $f_\mathrm{PBH}$ for different CR propagation models, assuming monochromatic mass functions, SUNE GMF model, and NFW DM profile. }
    \label{fig: fPBH-CR-SUNE_NFW}
\end{figure}

In Fig.~\ref{fig: fPBH-CR-SUNE_NFW}, we show the constraints on $f_\mathrm{PBH}$ for different CR propagation models, assuming monochromatic mass functions, SUNE GMF model, and NFW DM profile. 
For the DBz4 model which does not include the re-acceleration, we obtain modest constraints only for $M_c<1\times 10^{15}\,\mathrm{g}$, as can be expected from Fig.~\ref{fig: synch spec}. 
In contrast, for all diffusive re-acceleration models, we obtain stringent constraints. 
For instance, the most conservative constraints (obtained with the GH model) require $f_\mathrm{PBH} < 1.25 \times 10^{-4}$ for $M_c = 1.9 \times 10^{16}\,\mathrm{g}$. 
It can be seen that, as the diffusion halo half-height $z_\mathrm{h}$ increases from 4 kpc to 10 kpc (in the DRz4-10 models), the constraints strengthen by approximately an order of magnitude. 
For the GH and DRz4 models whose $z_\mathrm{h}=4\,\mathrm{kpc}$, the obtained constraints differ slightly for $M_c\gtrsim 5\times 10^{15}\,\mathrm{g}$. 

\begin{figure}[htbp]
    \centering
    \includegraphics[width=\textwidth]{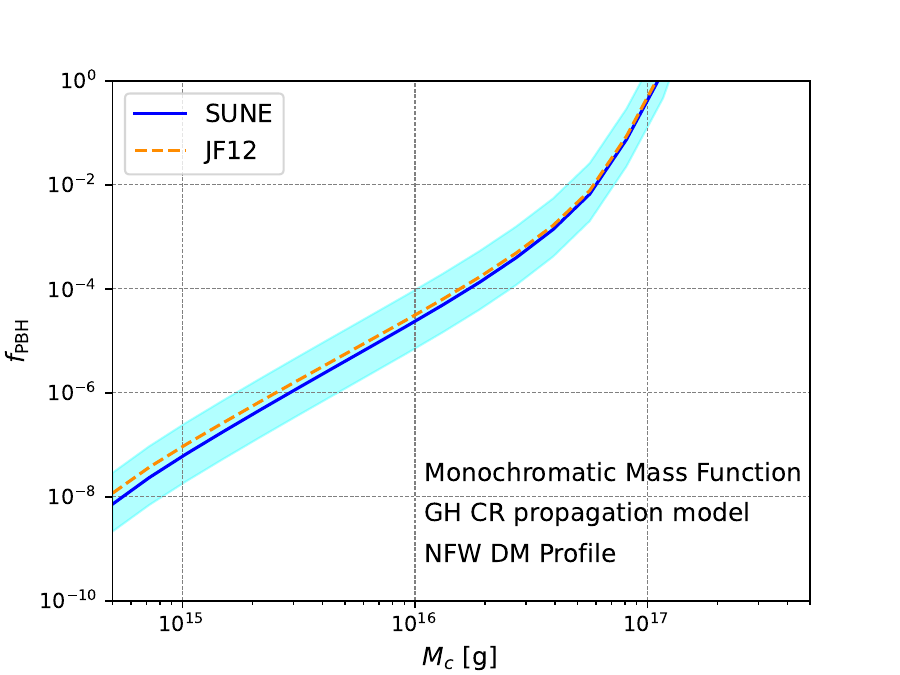}
    \caption{Constraints on $f_\mathrm{PBH}$ for SUNE and JF12 GMF models, assuming monochromatic mass functions, GH CR propagation model, and NFW DM profile. The shaded region illustrates the impact of varying the overall normalization of the SUNE magnetic field by a factor of two on the resulting constraints. }
    \label{fig: fPBH-Bfield}
\end{figure}

In Fig.~\ref{fig: fPBH-Bfield}, we show the constraints on $f_\mathrm{PBH}$ for SUNE and JF12 GMF models, assuming monochromatic mass functions, GH CR propagation model, and NFW DM profile. The shaded region illustrates the impact of varying the overall normalization of the SUNE magnetic field by a factor of two on the resulting constraints. 
It can be seen that, although these two GMF models have distinct geometrical structures and spatial distributions of magnetic field strength, the derived constraints are in remarkable agreement. Adopting alternative CR propagation models leads to similar results compared to the GH case. For completeness, we present these results in appendix~\ref{app: ex results} (see Fig.~\ref{fig: fPBH-GMF}).
To quantify the impact of magnetic field strength uncertainty, we vary the overall normalization of the SUNE magnetic field by a factor of two. This results in the range of constraints shown by the shaded region in Fig.~\ref{fig: fPBH-Bfield}. In particular, increasing the magnetic field strength by a factor of two strengthens the constraints by approximately a factor of 3.6. Nevertheless, the overall qualitative conclusions remain unchanged. 

Although the sky maps of the synchrotron signals of PBHs (see Fig.~\ref{fig: sky_DM_SUNE} and Fig.~\ref{fig: sky_DM_JF12}) for NFW and Burkert DM profiles look different, the constraints obtained with these two DM profiles differ slightly. 
We show the comparison of the constraints obtained with different DM profiles in Fig.~\ref{fig: fPBH-DM} in appendix \ref{app: ex results}. 
This agreement can be understood by that the main difference between the sky maps of synchrotron signals comes from the sky region near the Galactic center, where the Galactic synchrotron emission is too bright to set strong constraints. 
As will be shown later, the constraints are derived from the pixel located in dark areas in the observational sky map, and none of them is close to the Galactic center. 

We regard the constraints obtained with the GH model (the blue line in Fig.~\ref{fig: fPBH-CR-SUNE_NFW}) as our main conclusion. 
Because, firstly, the constraints are the most conservative among the results obtained with diffusive re-acceleration models. 
Secondly, the best-fit $z_\mathrm{h}=4.0\,\mathrm{kpc}$ of the GH model matches the CR electron densities used in the construction of both the SUNE and JF12 GMF models, improving consistency of our calculation. 

\begin{figure}[htbp]
    \centering
    \includegraphics[width=\textwidth]{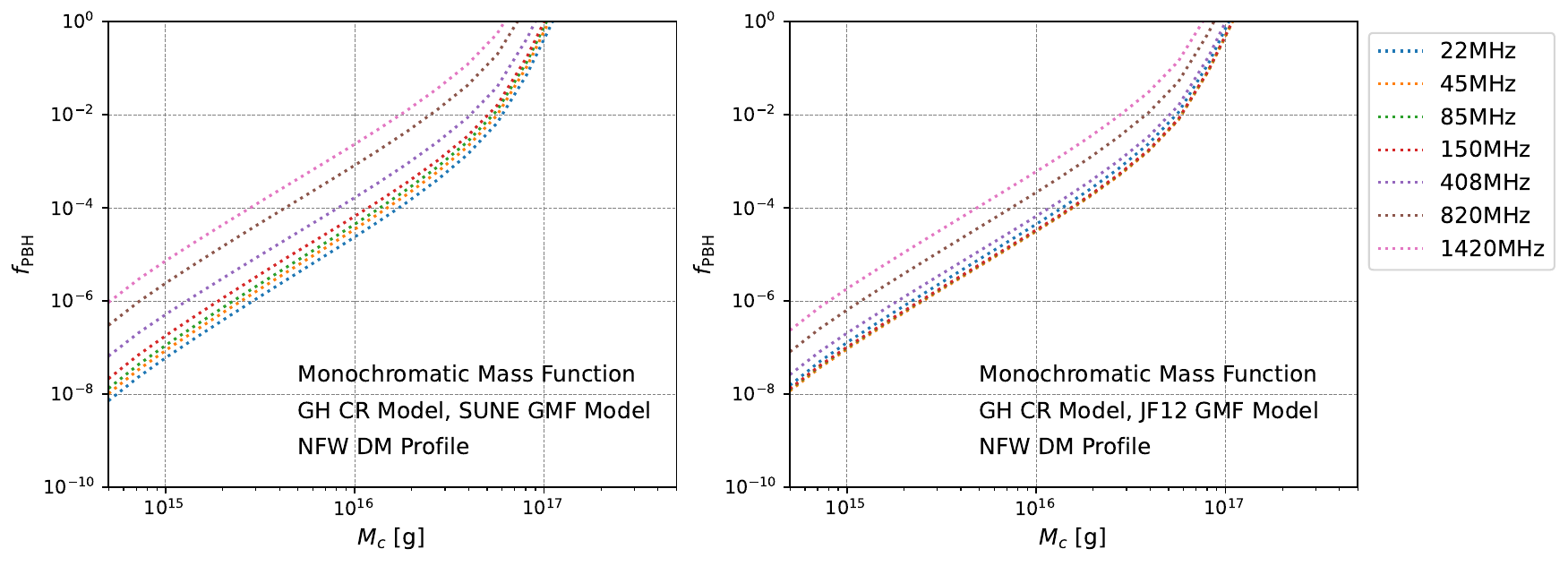}
    \caption{Constraints on $f_\mathrm{PBH}$ derived from individual observational sky maps, assuming monochromatic mass functions, GH CR propagation model, and NFW DM profile. Left: for SUNE GMF model. Right: for JF12 GMF model. }
    \label{fig: fPBH-freq}
\end{figure}

In Fig.~\ref{fig: fPBH-freq}, we show constraints on $f_\mathrm{PBH}$ derived from individual observational sky maps, assuming monochromatic mass functions, GH CR propagation model, NFW DM profile, and SUNE GMF model in the left, JF12 GMF model in the right, respectively. 
For the SUNE model, the constraints weaken as the observational frequency increasing, and the strongest constraints are derived from the 22 MHz observational sky map. 
For the JF12 model, the observational sky maps at frequencies from 45 MHz to 150 MHz set equal and the strongest constraints, while the constraints derived from the 22 MHz observational sky map are weaker slightly. 
Besides, the moderately weaker constraints derived from the 408 MHz observational sky map are particularly robust. 
Because the 408 MHz observational sky map is minimally contaminated by non-synchrotron processes, and the observational sky maps at frequencies from 22 MHz to 85 MHz are not full-sky. 

\begin{figure}[htbp]
    \centering
    \includegraphics[width=\textwidth]{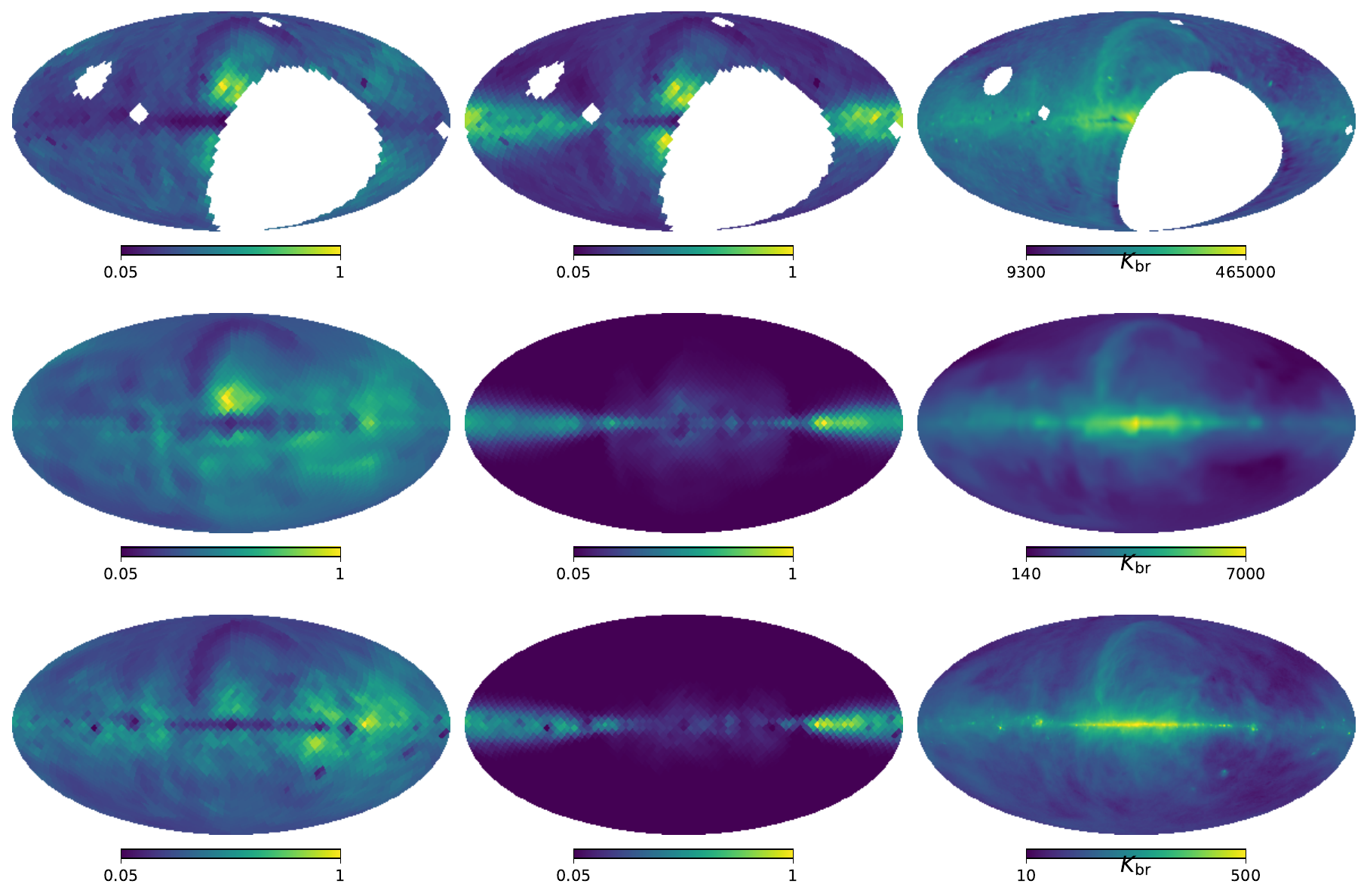}
    \caption{The first two columns: ratio of synchrotron signals to observational limits (intensity + $2\times$error) at 22, 150 and 408 MHz (top to bottom) for SUNE and JF12 GMF models, respectively, assuming a monochromatic mass function with $M_c=5\times 10^{14}\,\mathrm{g}$, GH CR propagation model, and NFW DM profile. Synchrotron signals are scaled by specific values of the constraints on $f_\mathrm{PBH}$ derived from individual observational sky maps such that ratio = 1 corresponds to the pixel where the constraint is obtained. All maps are plotted with $N_\mathrm{side}=16$ and linear color mapping. The last column: the 22, 150 and 408 MHz (top to bottom) observational sky maps for comparison, plotted with original resolution and logarithmic color mapping. }
    \label{fig: sky_ratio_NFW}
\end{figure}

In Fig.~\ref{fig: sky_ratio_NFW}, we show the ratio of synchrotron signals to observational limits (intensity + $2\times$error) at 22, 150 and 408 MHz (top to bottom) for SUNE and JF12 GMF models, respectively, assuming a monochromatic mass function with $M_c=5\times 10^{14}\,\mathrm{g}$, GH CR propagation model, and NFW DM profile. Synchrotron signals are scaled by specific values of the constraints on $f_\mathrm{PBH}$ derived from individual observational sky maps such that ratio~=~1 corresponds to the pixel where the constraint is obtained. 
Observational sky maps at corresponding frequencies are also shown for comparison. 
Results for the Burkert profile (see Fig.~\ref{fig: sky_ratio_Bur}) are put in the appendix \ref{app: ex results}, which do not largely differ from those for the NFW profile. 
From the figures, it can be seen that the Galactic center is too bright to set strong constraints. Instead, the most constraining sky regions are: 
\begin{itemize}
    \item \textbf{For the SUNE model}: in the 22 MHz ratio map, the region $\sim 30^\circ$ above the Galactic center within $\sim 10^\circ$ radius, corresponding to the dark area enclosed by the North Polar Spur in the observational sky map. 
    \item \textbf{For the JF12 model}: in the 150 MHz ratio map, the region $\sim 15^\circ$ below and above the Galactic disc at longitudes $240^\circ<l<300^\circ$, corresponding to the dark area on the Galactic disc at the same longitudes in the observational sky map. 
\end{itemize}
Note that in the right panel of Fig.~\ref{fig: fPBH-freq}, the constraints derived from the 22 MHz observational sky map for the JF12 model are slightly weaker than the strongest constraints, because the outer-disc region (at longitudes $240^\circ<l<300^\circ$) absents in the 22 MHz observational sky map. 
The difference between the most constraining sky regions for these two GMF models likely origins from JF12's prominent disc random field (especially in 6th and 7th spiral arms), which yields larger synchrotron signals on the Galactic disc. 

\begin{figure}[htbp]
    \centering
    \includegraphics[width=\textwidth]{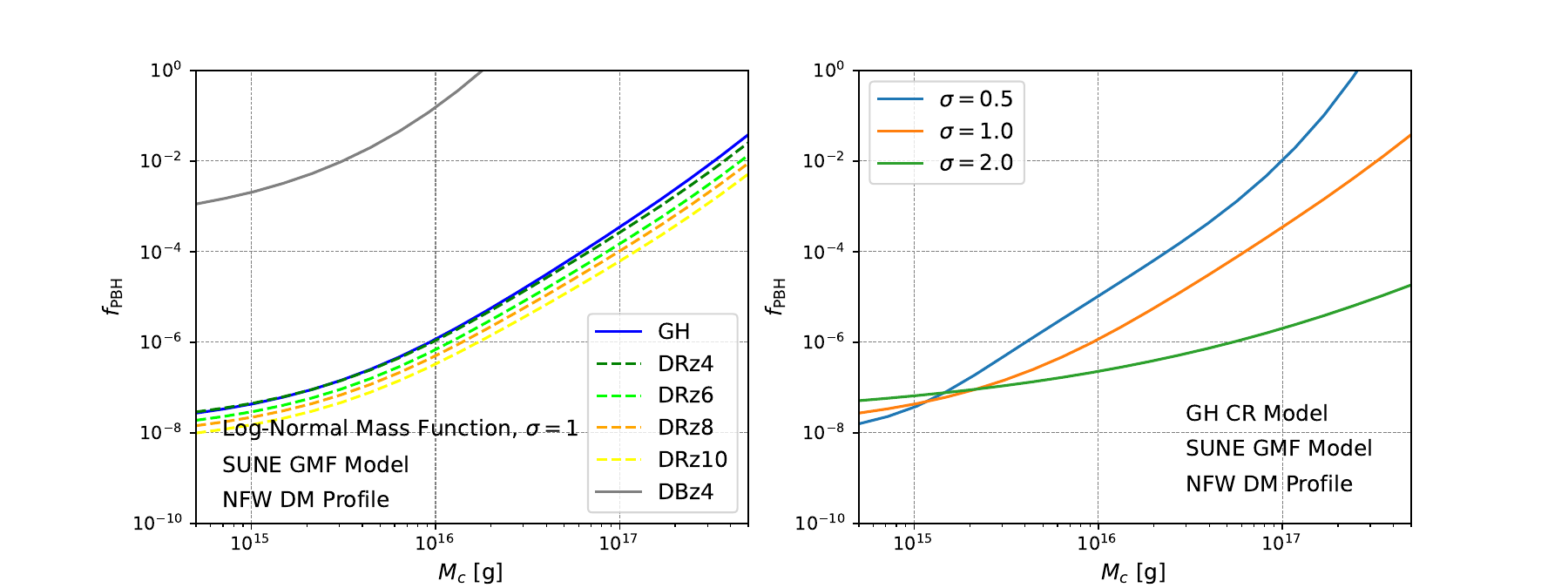}
    \caption{Constraints on $f_\mathrm{PBH}$ with log-normal mass functions. Left: for different CR propagation models, assuming the distribution width $\sigma=1$, SUNE GMF model, and NFW DM profile. Right: for different distribution widths $\sigma=0.5$, 1.0, and 2.0, assuming GH CR propagation model, SUNE GMF model, and NFW DM profile.}
    \label{fig: fPBH_logn}
\end{figure}

We now extend our analysis to log-normal PBH mass functions, which are often considered more physically motivated in formation scenarios. 
In Fig.~\ref{fig: fPBH_logn}, we show constraints on $f_\mathrm{PBH}$ with log-normal mass functions. The left panel of Fig~\ref{fig: fPBH_logn} shows results for different CR propagation models, assuming the distribution width $\sigma=1$, SUNE GMF model, and NFW DM profile. 
Similar to the monochromatic case, we obtain stringent constraints with diffusive re-acceleration models, while the constraints obtained with the diffusion break model are much weaker. Adopting a log-normal mass function modifies the mass dependence of the constraints compared to the monochromatic case. While the constraints at the lowest masses (e.g., $\sim 10^{15}\,\mathrm{g}$) remain nearly unchanged (compared with the lines in Fig~\ref{fig: fPBH-CR-SUNE_NFW}), the limits derived with the log-normal mass function weaken more slowly with increasing $M_c$. As a result, around $10^{16}\,\mathrm{g}$ the constraints become stronger than the limits of monochromatic case by approximately one order of magnitude for $\sigma = 1$. This behavior arises because the extended mass function retains a significant contribution from lighter PBHs, which emit more and higher-energy all-electrons via Hawking radiation, resulting in stronger synchrotron emission. While for the lowest masses, this effect vanishes as we impose a lower bound of the PBH mass ($M_\mathrm{min}=5\times 10^{14}\,\mathrm{g}$). 
The right panel of Fig~\ref{fig: fPBH_logn} compares the constraints across distribution widths $\sigma=0.5$, 1.0, and 2.0, assuming GH CR propagation model and otherwise identical conditions. 
Constraints strengthen with increasing $\sigma$ for $M_c>5\times 10^{15}\,\mathrm{g}$, because broader mass functions (at fixed $M_c$) retain lighter PBHs, whose larger contribution to synchrotron signals effectively strengthen the constraints. 
Since the PBH mass function affects the injection spectrum of evaporated all-electrons, rather than their spatial dependence, the sky maps of synchrotron signals remain similar across $\sigma$ values. 
The effects of changing different GMF models and DM profiles are similar to those in the monochromatic case, which can be regard as a log-normal limit as $\sigma\rightarrow 0$. So we omit redundant exhibitions. 

\begin{figure}[htbp]
    \centering
    \includegraphics[width=\textwidth]{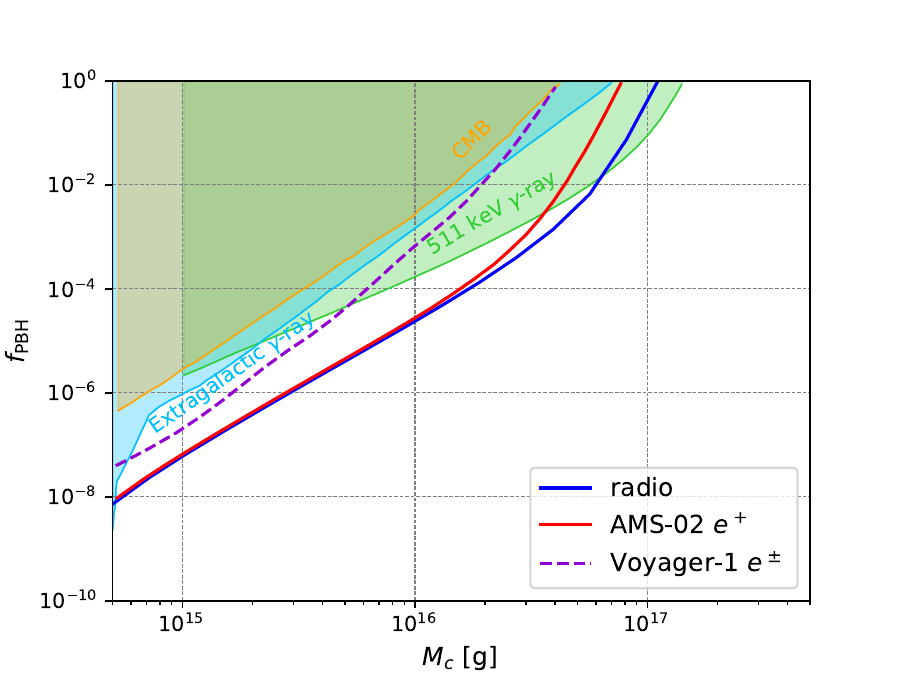}
    \caption{Several constraints on $f_\mathrm{PBH}$ with monochromatic mass functions. 
    Shadow regions show constraints derived from different observables, including extragalactic $\gamma$-rays \cite{Carr:2009jm}, CMB \cite{Auffinger:2022khh}, and 511 keV $\gamma$-rays \cite{Keith:2021guq}. 
    Lines show constraints derived from Voyager-1 all-electrons data \cite{Boudaud:2018hqb}, AMS-02 positron data \cite{Huang:2024xap}, and radio continuum surveys (assuming GH CR propagation model, SUNE GMF model, and NFW DM profile). }
    \label{fig: fPBH comp}
\end{figure}

In Fig.~\ref{fig: fPBH comp}, we compare our constraints on $f_\mathrm{PBH}$ derived from radio continuum surveys with monochromatic mass functions (assuming GH CR propagation model, SUNE GMF model, and NFW DM profile) with several existing limits. These include constraints derived from the data of extragalactic $\gamma$-rays \cite{Carr:2009jm}, CMB \cite{Auffinger:2022khh}, 511 keV $\gamma$-rays \cite{Keith:2021guq}, Voyager-1 all-electrons \cite{Boudaud:2018hqb}, and AMS-02 positrons \cite{Huang:2024xap}. 
The result from \cite{Keith:2021guq} was obtained assuming an NFW inner slope of $\gamma=1.6$, making it rather conservative. While \cite{DelaTorreLuque:2024qms} obtained much stronger constraints from the 511 keV $\gamma$-ray data. 
The results from \cite{Boudaud:2018hqb} and \cite{Huang:2024xap} were obtained without including the astrophysical backgrounds, and using the same method to calculate the constraints on $f_\mathrm{PBH}$ as in this work, i.e., requiring that the predicted signal does not exceed the data by more than twice the experimental uncertainty. 
Thus, we can compare our constraints with them directly. 
Moreover, our previous constraints derived from AMS-02 positron data \cite{Huang:2024xap} were obtained using the same GH CR propagation model. 

From Fig.~\ref{fig: fPBH comp}, it can be seen that the constraints derived from radio continuum surveys are stronger than those derived from the Voyager-1 all-electron data \cite{Boudaud:2018hqb} by more than one order of magnitude for $M_\mathrm{PBH} \gtrsim  1\times 10^{16}\,\mathrm{g}$, and also stronger than our previous constraints derived from AMS-02 positron data \cite{Huang:2024xap} for $M_\mathrm{PBH} \gtrsim 2\times 10^{16}\,\mathrm{g}$. 
The CR positrons are believed to be of secondary origin and relatively rare, making them sensitive to exotic contributions. 
It is therefore noteworthy that radio continuum surveys can set constraints that are comparable to, or even stronger than, those derived from AMS-02 positron data. 
The CR particles lose energy when traveling through the heliosphere. 
Based on the force-field approximation, interstellar positrons at least need a kinetic energy $E_\mathrm{k} \gtrsim E_\mathrm{AMS,min} + e\phi \approx 0.57\,\mathrm{GeV} + 0.5\,\mathrm{GeV} \approx 1\,\mathrm{GeV}$ to be detected by AMS-02, where $E_\mathrm{AMS,min}=0.57\,\mathrm{GeV}$ is the kinetic energy of the lowest-energy AMS-02 positron data point, and $\phi=500\,\mathrm{MV}$ is a conservatively estimated modulation potential. 
In contrast, the observation of Galactic synchrotron emissions at frequencies of $\sim20\,\mathrm{MHz}$ can provide indirect measurements of interstellar CR all-electrons with energies of $\sim 100\,\mathrm{MeV}$, and therefore can set stronger constraints on heavier PBHs, for which the energy spectra of evaporated all-electrons are shifted to lower energies. 
For the heaviest PBHs with masses $M_\mathrm{PBH}\gtrsim 1\times10^{17}\,\mathrm{g}$, their temperatures become too low to emit significant all-electrons, causing all constraints to weaken. 
Furthermore, the synchrotron emissions can travel though the Galaxy, thus our constraints can complement those derived from local CR measurements. 

It is worth noting what fundamentally limits our current conservative constraints and the implications for future radio probes. 
At low radio frequencies, the instrumental thermal noise is negligible compared to the extremely bright Galactic synchrotron emission. 
Therefore, our bounds are not primarily limited by instrumental thermal noise, but rather by the intensity of the astrophysical radio background and its spatial variance (through the observational error). 
The next-generation radio facilities, such as the SKA-low and improved LOFAR surveys, are expected to meaningfully tighten these constraints. 
They will achieve this through their unprecedented angular resolution, which can resolve and remove extragalactic point sources, reducing the small-scale spatial variance in the observational maps. 
Furthermore, their advanced broad-band and polarimetric capabilities will make it possible to construct a robust GMF model and subtract the astrophysical radio background. 
By constraining $f_\mathrm{PBH}$ using only the much smaller residual intensity rather than the total observed intensity, future limits could be improved by an order of magnitude. 
On the other hand, since the predicted spectra of synchrotron signals of PBHs are softer than the observed spectrum, future radio surveys extending below the atmospheric window threshold (10 MHz) could provide stronger constraints. It is also possible to obtain meaningful constraints on $f_\mathrm{PBH}$ with diffusion break models of CR propagation, if the radio survey data at lower frequencies are available.

\section{Conclusion}   \label{sec: conclusion}
In this work, we have investigated the possibility of constraining PBHs with masses $M_\mathrm{PBH}\gtrsim 10^{15}\,\mathrm{g}$ through Galactic diffuse synchrotron emissions. 
Using the AMS-02 \cite{AMS:2023anq} and Voyager-1 \cite{Cummings:2016pdr} data on the B/C flux ratio, we fit parameters of a set of benchmark CR propagation models, and confirm that a significant Alfv\'{e}n velocity $V_a \sim 20\,\mathrm{km/s}$ is favored in several diffusive re-acceleration models. 
We show that, for $V_a\sim 20\,\mathrm{km/s}$, a significant fraction of the evaporated all-electrons, which typically have initial energies of $\sim 10\,\mathrm{MeV}$, can be boosted to the energies of $\sim 100\,\mathrm{MeV}$ during their Galactic propagation. 
Consequently, synchrotron emissions generated by the evaporated all-electrons can be constrained by the low-frequency radio continuum surveys (from 22 MHz to 1.4 GHz). 
With diffusive re-acceleration models, we obtain stringent constraints on $f_\mathrm{PBH}$. 
The most conservative constraints are stronger than those derived from the Voyager-1 all-electron data \cite{Boudaud:2018hqb} by more than one order of magnitude for $M_\mathrm{PBH} \gtrsim  1\times 10^{16}\,\mathrm{g}$, and also stronger than our previous constraints derived from the AMS-02 positron data \cite{Huang:2024xap} for $M_\mathrm{PBH} \gtrsim 2\times 10^{16}\,\mathrm{g}$. 
We have checked the robustness of our constraints against different DM profiles and uncertainties in the GMF model, and we have also extended our analysis to log-normal PBH mass functions. 

Admittedly, the main results of this work critically depend on the diffusive re-acceleration models of CR propagation, which are supported by a number of independent analyses \cite{Trotta:2010mx, Jin:2014ica, Johannesson:2016rlh, Boschini:2017fxq, Boschini:2018baj, Boschini:2020jty, DeLaTorreLuque:2021yfq, Luque:2021nxb, Yuan:2017ozr, Korsmeier:2021brc, Silver:2024ero}, but not yet conclusive. 
Future X-ray observation missions (e.g., e-ASTROGAM, AMEGO \cite{Engel:2022bgx}) can probe interstellar all-electrons with energies of $\mathcal{O}(100\,\mathrm{MeV})$ via their inverse Compton effects, potentially examining the existence of diffusive re-acceleration \cite{Orlando:2017mvd}. 
Finally, we emphasize that our present bounds are primarily limited by the intensity and spatial variance of the astrophysical radio background. 
Future low-frequency radio facilities such as SKA-low and improved LOFAR surveys with improved angular resolution and broad-band polarimetric capabilities are expected to enable more robust modeling of the GMF and subtraction of the astrophysical radio background in the analysis. 
Such improvements could tighten the constraints on $f_\mathrm{PBH}$ by an order of magnitude. 
In addition, since the spectra of synchrotron signals of PBHs are softer than the observed spectrum, future radio surveys extending to frequencies below the atmospheric window threshold can also provide stronger constraints on PBHs.

\begin{acknowledgments}
This work is supported in part by the NSFC under Grants No. 12441504 and No. 12447101. 
\end{acknowledgments}

\bibliographystyle{arxivref.bst}
\bibliography{PBH}

\newpage

\appendix
\section{Fit Results for Benchmark Models}   \label{app: CR}
We fit parameters of a set of benchmark CR propagation models using the AMS-02 \cite{AMS:2023anq} and Voyager-1 \cite{Cummings:2016pdr} data on the B/C flux ratio. 
Our model set includes four diffusive re-acceleration models (DRz4-10 models) with different diffusion halo half-heights $z_\mathrm{h}$ (fixed to 4, 6, 8, 10 kpc, respectively), and one diffusion break model (DBz4 model) with $z_\mathrm{h}$ fixed to 4 kpc. 
In section~\ref{sec: benchmark CR} of the main text, we have introduced the set of model free parameters and the fitting procedure, and shown the the estimated model parameters (see Tab.~\ref{tab: CR_best}) and the B/C flux ratio predictions (see Fig.~\ref{fig: bestfit_BC}) for the benchmark models. 
In this section, we describe the numerical details of our fits, show the C flux predictions for the benchmark models, and display corner plots (showing the joint and marginal posterior distributions) of the posterior PDFs for the benchmark diffusive re-acceleration models. 

\subsection{Numerical Details} 
The \texttt{Galprop} configurations are fixed as follows: We employ 2D spatial grids assuming cylindrical symmetry of the Galaxy. The diffusion halo extends radially to $R=20\,\mathrm{kpc}$ and vertically to $z=\pm z_\mathrm{h}$, with $z_\mathrm{h}$ values of 4, 6, 8, and 10 kpc for the respective models. We set linear spatial grids with 41 vertical grid points and 24 radial grid points, the latter one is adjusted automatically by \texttt{Galprop} because we adopt the vectorized Crank-Nicolson solution method (with GALDEF option \texttt{solution\_method = 4}). The spatial dependence of the CR primary source follows Eq.~\eqref{eq: CR spatial}, with $a=1.9$, $b=5.0$ \cite{Lorimer:2006qs}, and $z_0=0.2\,\mathrm{kpc}$. Primary isotopic abundances, except that of $^{12}\mathrm{C}$, are fixed to standard \texttt{Galprop} values \citep{Moskalenko:2007ig}. The proton flux normalization at 100 GeV is $4.49\times 10^{-2}\,\mathrm{GeV^{-1}m^{-2}s^{-1}sr^{-1}}$, which is used to determine the normalization of primary CR source term. The nuclear reaction network includes elements up to Si (with charge number $Z=14$), utilizing the cross section parameterization corresponding to the GALDEF option \texttt{kopt=012}. For interstellar gas components, we adopt the standard \texttt{Galprop} released CO and H\,I distributions, and the H\,II model as in \cite{Orlando:2013ysa}. We use the \texttt{MultiNest} package \cite{Feroz:2007kg, Feroz:2008xx, Feroz:2013hea} to sample posterior PDFs, of which the dimension is 9 for both types of models. For the \texttt{MultiNest} configurations, we set 400 live points, evidence tolerance of 0.5 and sampling efficiency of 0.8. 

\subsection{Extended Results}

\begin{figure}[htbp]
    \centering
    \includegraphics[width=\textwidth]{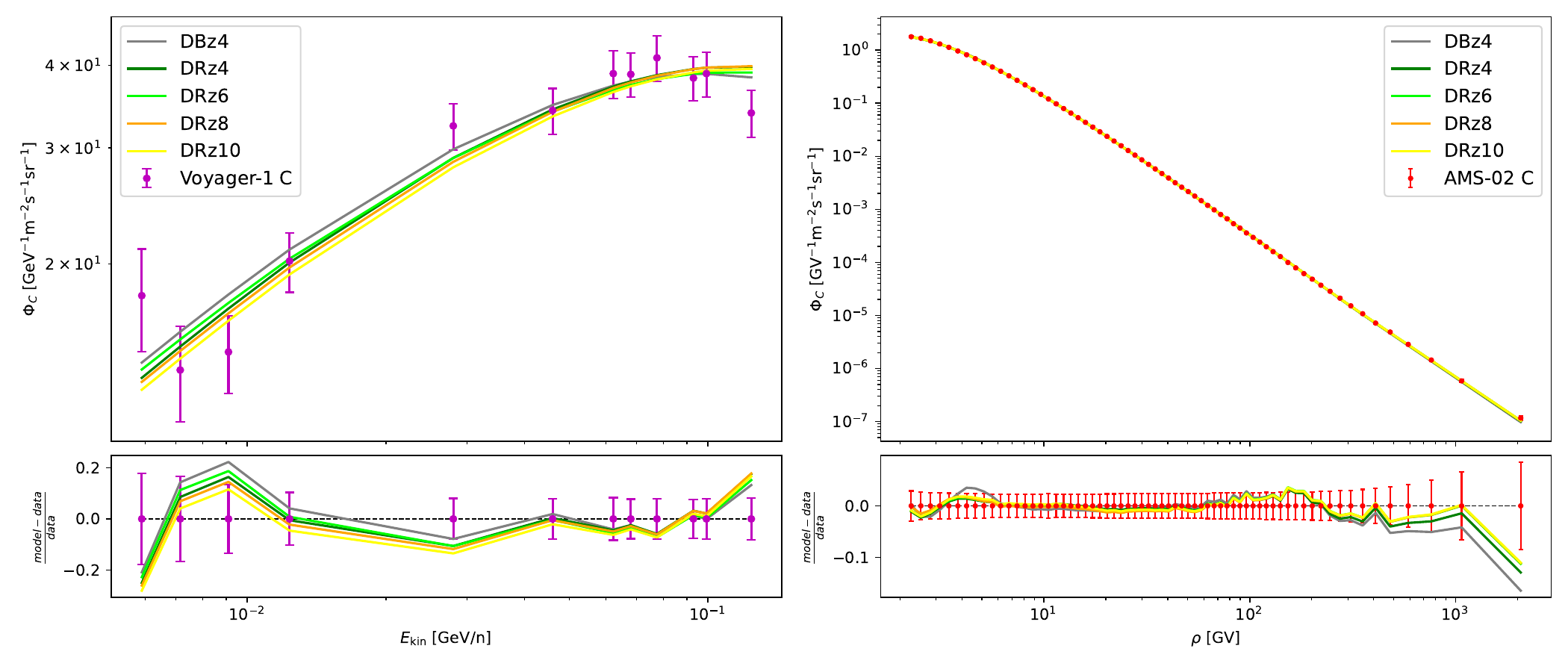}
    \caption{Model predictions and residuals of C flux for the benchmark models. Left: for Voyager-1. Right: for AMS-02.}
    \label{fig: bestfit_C}
\end{figure}

\begin{figure}[htbp]
    \centering
    \includegraphics[width=\textwidth]{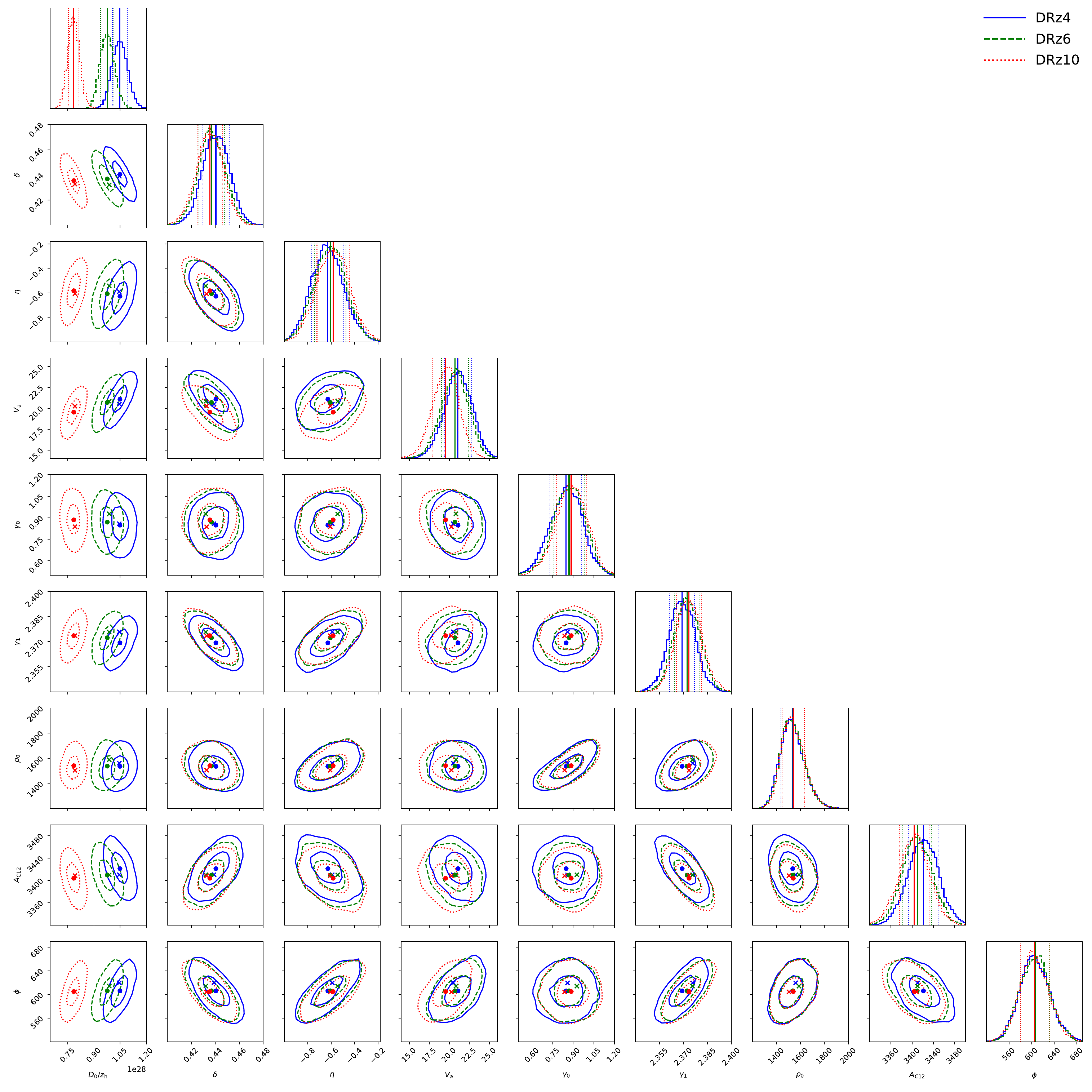}
    \caption{Corner plots of posterior PDFs for DRz4, DRz6, and DRz10 models. The result of DRz8 model is excluded for clarity. Contours show 1 and 2-$\sigma$ credible regions of 2D distributions, containing 39.3\% and 86.4\% of the posterior probability mass, respectively. Points mark the posterior means, and crosses mark the maximum posterior values. In 1D marginalized posterior plots, solid vertical lines show the posterior means, and dotted vertical lines show positions of the posterior mean $\pm$ standard deviation. For a tight plot, $D_0$ is replaced with $D_0/z_\mathrm{h}$ (with unit: $10^{28}\, \mathrm{cm}^2\, \mathrm{s}^{-1}\, \mathrm{kpc}^{-1}$). }
    \label{fig: triangle}
\end{figure}

In Fig.~\ref{fig: bestfit_C}, we show model predictions and residuals of C flux for the benchmark models. It can be seen that all DRz4-10 models provide equally good fits to Voyager-1 and AMS-02 data, yielding nearly identical predictions. While the DBz4 model exhibits slightly insufficient spectral hardening around 300 GV. In Fig.~\ref{fig: triangle}, we show corner plots of posterior PDFs for DRz4, DRz6, and DRz10 models. The result of DRz8 model is not shown for clarity. The corner plots for the DBz4 model are omitted, since our main conclusion does not rely on this model. For a tight plot, $D_0$ is replaced with $D_0/z_\mathrm{h}$ (with unit: $10^{28}\, \mathrm{cm}^2\, \mathrm{s}^{-1}\, \mathrm{kpc}^{-1}$). It can be seen that all parameters, except for $D_0/z_\mathrm{h}$, remain approximately constant across $z_\mathrm{h}$. As for $D_0/z_\mathrm{h}$, it decreases slightly as the $z_\mathrm{h}$ increasing. This feature is also obtained by other works, for example, see \cite{Johannesson:2016rlh}. 

\section{Extended Results for Constraints on the PBH Abundance}   \label{app: ex results}
This appendix serves as supplementary materials of section~\ref{sec: fPBH constraints}. Below, we show several results of constraints on $f_\mathrm{PBH}$ that are of less importance or redundant to be shown in the main text. 

\begin{figure}[htbp]
    \centering
    \includegraphics[width=\textwidth]{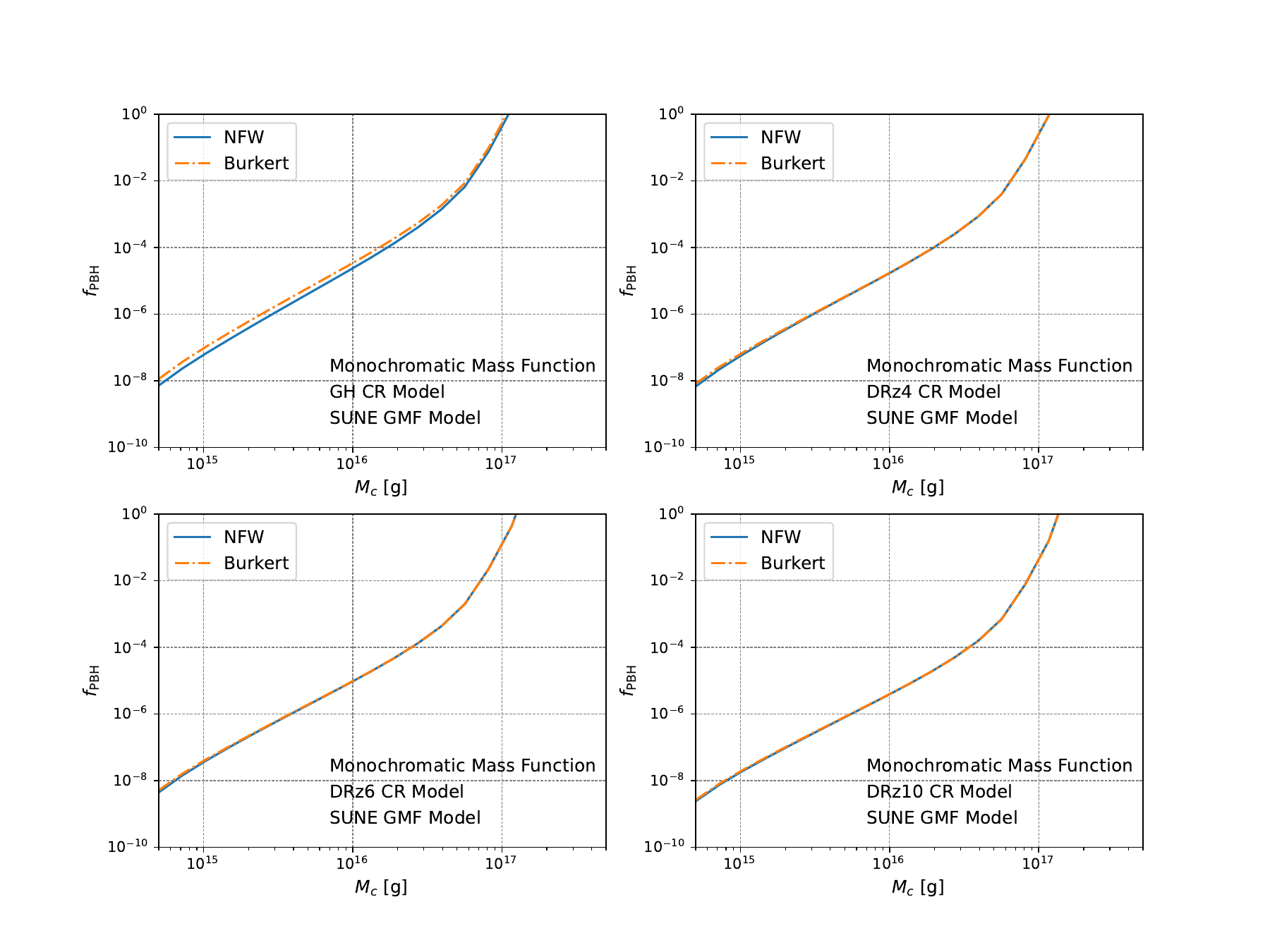}
    \caption{Constraints on $f_\mathrm{PBH}$ for different DM profiles and for different CR propagation models (as labeled), assuming monochromatic mass functions and SUNE GMF model. The plot for DRz8 model is omitted.}
    \label{fig: fPBH-DM}
\end{figure}

In Fig.~\ref{fig: fPBH-DM}, we show constraints on $f_\mathrm{PBH}$ for different DM profiles and for different CR propagation models (as labeled in the figure), assuming monochromatic mass functions and SUNE GMF model. The plot for DRz8 model is omitted. 
It can be seen that the constraints obtained with NFW and Burkert profiles are nearly equal for all diffusive re-acceleration models. 
While for JF12 GMF model, the difference between the constraints obtained with these two DM profiles are less obvious, thus the results are not shown. 

\begin{figure}[htbp]
    \centering
    \includegraphics[width=\textwidth]{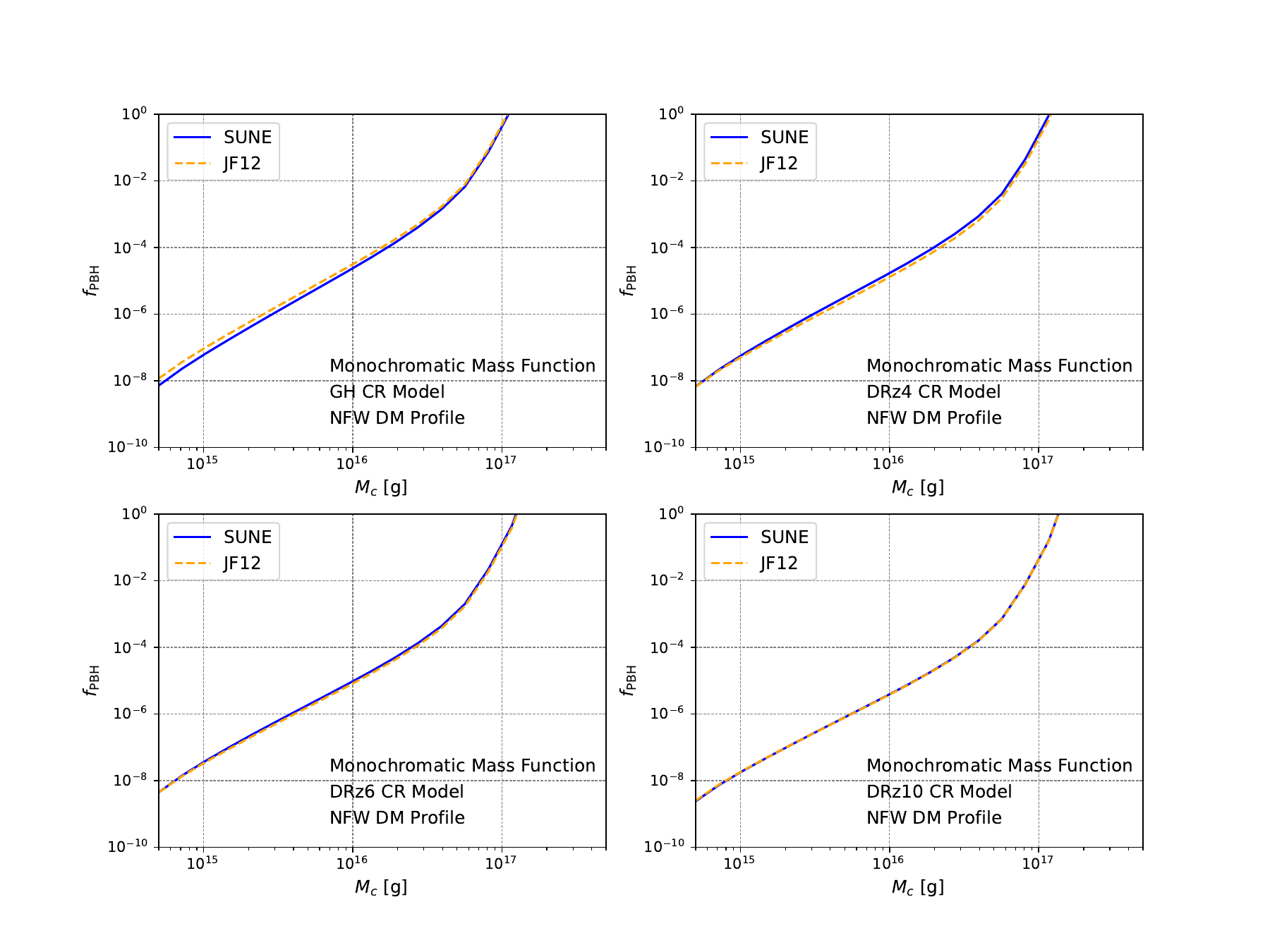}
    \caption{Constraints on $f_\mathrm{PBH}$ for different GMF models and for different CR propagation models (as labeled), assuming monochromatic mass functions and NFW DM profile. The plot for DRz8 model is omitted.}
    \label{fig: fPBH-GMF}
\end{figure}

In Fig.~\ref{fig: fPBH-GMF}, we show constraints on $f_\mathrm{PBH}$ for different GMF models and for different CR propagation models (as labeled in the figure), assuming monochromatic mass functions and NFW DM profile. The plot for DRz8 model is omitted. 
It can be seen that the constraints obtained with SUNE and JF12 GMF models are nearly equal for all diffusive re-acceleration models. 
The results for Burkert DM profile are quite similar, thus are not shown. 

\begin{figure}[htbp]
    \centering
    \includegraphics[width=\textwidth]{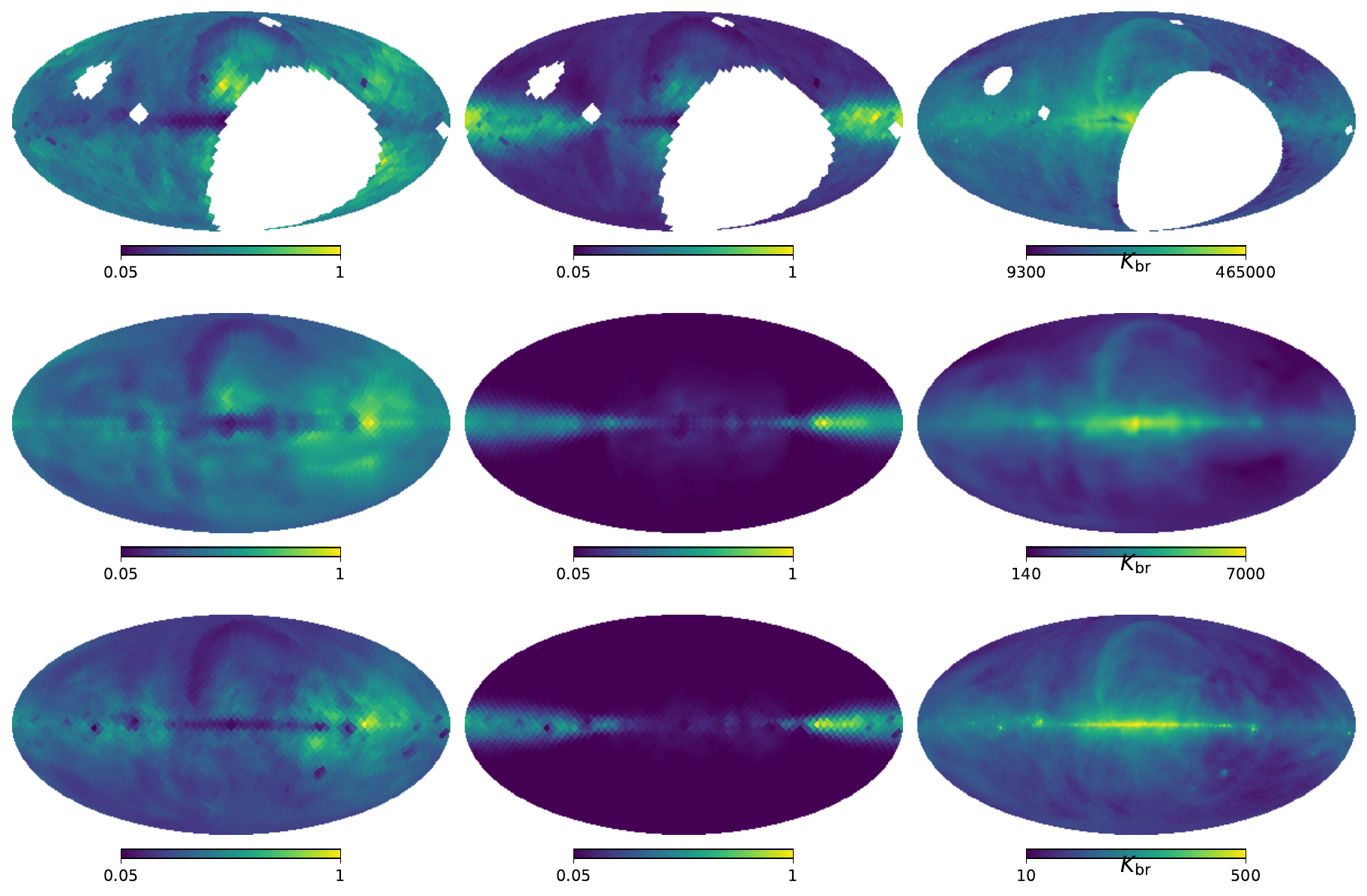}
    \caption{Same as Fig.~\ref{fig: sky_ratio_NFW} but for Burkert DM profile. }
    \label{fig: sky_ratio_Bur}
\end{figure}

In Fig.~\ref{fig: sky_ratio_Bur}, we show the ratio of synchrotron signals to observational limits (intensity + $2\times$error) at 22, 150 and 408 MHz (top to bottom) for SUNE and JF12 GMF models, respectively, assuming the monochromatic mass function with $M_c=5\times 10^{14}\,\mathrm{g}$, GH CR propagation model, and Burkert DM profile. 
Synchrotron signals are scaled by specific values of the constraints on $f_\mathrm{PBH}$ derived from individual observational sky maps such that ratio~=~1 corresponds to the pixel where the constraint is obtained. 
Observational sky maps at corresponding frequencies are also shown for comparison. 
The most constraining sky regions are similar to those for NFW profile, while the outer-disc region becomes more prominent, especially for the SUNE model at frequencies of $\gtrsim 150\,\mathrm{MHz}$. 
However, the most constraining sky region in the 22 MHz sky map for the SUNE model still matches the dark area enclosed by the North Polar Spur, as the outer-disc region absents in the 22 MHz observational sky map. 
\end{document}